\documentclass[10pt]{iopart}\usepackage{iopams}  
\usepackage{graphicx}
\usepackage{mathrsfs} 
\usepackage{fancyhdr} 

\usepackage{epsfig}
\usepackage{eucal} 
\usepackage{amssymb}
\usepackage{amsfonts}
\usepackage{amsbsy}

\begin{document}

\title{Continuous Emission of A Radiation Quantum}

\author{J.X. Zheng-Johansson
}
\address{
Institute of Fundamental Physics Research, 611 93 Nyk\"oping, Sweden  
} 

\def\citejxzjied{3}
\def\xbar{{\bar{x\hspace{0.1cm}}\hspace{-0.1cm}}}
\def\Vo{\overline{V}}
\def\Ho{\overline{H}}

\def\Isub{{\mbox{\tiny${I}$}}}
\def\la{\langle}
\def\ra{\rangle}
\def\Vol{{V\hspace{-0.29cm}^{_{\mbox{-}}} \hspace{0.13cm}}}
\def\Voll{{V\hspace{-0.29cm}^{_{\mbox{-}}} \hspace{0.13cm}}}
 \def\Vols{{V\hspace{-0.24cm}^{_{\mbox{-}}} \hspace{0.13cm}}}
\def\pa{{''}}
\def\No{{N}}
\def\psip{\psi_{r,q}{}}
\def\pspn{\psi_{r,n_q}{}}
\def\xib{{\pmb{\xi}}}
\def\Mv{\mathcal{M}}
\def\Aqb{{\bf{A}}}

\def\Vscr{V}

\def\Pb{{\bf{P}}}
\def\Auv{{\Aqb\hspace{-0.2cm}{{}^{^{^{_o}}}}}} 
\def\ruv{{\rb\hspace{-0.15cm}{{}^{^o}}}}
\def\ka{\kappa}

\def\to{t_o}
\def\tomu{{}}

\def\taut{t}

\def\timu{{t_{i_\mus}}}
\def\tfmu{{t_{f_\mus}}}
\def\teqmu{{t_{o \mus}}}
\def\aimu{{a_{1_\mus}}}
\def\aiimu{{a_{2_\mus}}}
\def\taumu{{\taut_{n_\mus,n_\mus'}}}
\def\mus{{}}
\def\nus{{}}
        \def\musp{\mu}
\def\nusp{\nu}

\def\ai{a}
\def\aii{{a_r}}

\def\xb{{\bf{x}}}
\def\zb{{\bf{z}}}
\def\Xb{{\bf{X}}}
\def\Zb{{\bf{Z}}}


\def\Aqo{A}
\def\Efrak{{\Ef'}}
\def\Eftotb{{\Efb}}
\def\Emb{{{\mbox{\sf{E}}}}}
\def\Ef{{{\mbox{\sf{E}}}}}

\def\Bf{{{\mbox{\sf{B}}}}}
\def\Bmb{{{\mbox{\sf{B}}}}}
\def\Embb{{{\pmb{\mbox{\sf{E}}}}}}
\def\Bmbb{{{\pmb{\mbox{\sf{B}}}}}}
\def\Efb{{{\pmb{\mbox{\sf{E}}}}}}
\def\Bfb{{{\pmb{\mbox{\sf{B}}}}}}

\def\Aqb{{\bf{A}}}

\def\Ab{{{\pmb{\mbox{\sf{A}}}}}}
\def\Ao{{{\mbox{\sf{A}}}}}
\def\Asf{{{\mbox{\sf{A}}}}}
\def\Avb{{{\pmb{\mbox{\sf{A}}}}}}
\def\Afb{{{\pmb{\mbox{\sf{A}}}}}}
\def\Avo{{{\mbox{\sf{A}}}}}
\def\Yfb{{{\pmb{\mbox{\sf{Y}}}}}}
\def\Av{{{\mbox{\sf{A}}}}}
\def\Af{{{\mbox{\sf{A}}}}}

\def\Re{{\rm Re}}
\def\si{{s_1}}
\def\Ycal{{\mathcal Y}}
\def\f{\mathcal{Y}}
\def\Asub{{\mbox{\tiny${\Ycal}$}}}

\def\rw{\rightarrow}
\def\Hcal{{\mathcal{H}}}
\def\Hscr{{\mathscr{H}}}
\def\engvel{\eng_d{}}
\def\engvela{\eng'_d{}}
\def\engvelt{\eng'_d{}}

\def\engqkp{{\mathscr{K}}'}
\def\engrk{{\mathcal{K}}}
      \def\engqk{\eng_{q.kin}{}}

\def\engqkvel{\engt_{q\vel.kin}}
\def\Psub{{\mbox{\tiny${P}$}}}
\def\Ptsub{\tilde{\Psub}}
\def\Ptsub{{\mbox{\scriptsize{${\mathscr{P}}$}}}}
\def\Pt{{\mbox{${\mathscr{P}}$}}}
\def\engtsub{{\mbox{\scriptsize{${\mathscr{E}}$}}}}
\def\engtsup{{\mbox{\scriptsize{${\mathscr{E}}$}}}}
\def\Mt{\widetilde{M}}
\def\Engt{\widetilde{\Eng}}
\def\Engtsub{\tilde{\Eng}}
\def\mt{\tilde{m}}

\def\T{U}

\def\engqvelt{\engt_d{}}        
\def\engqvel{\eng_d{}}

\def\Tsub{{\mbox{\tiny${\T}$}}}
\def\engveltsub{\tilde{\Tsub}}        

\def\engtotvelt{\widetilde{\T}_{tot}{}}
\def\engtotvel{\T_{tot}{}}

\def\engk{\engt_{kin}}

\def\wo{\w}
\def\Wo{\W}
\def\wp{\w^{\dagsup}}

\def\wm{{\w_d}}

\def\sig{\sigma}
\def\engden{\rho_{eng}}
\def\qrk{{qrk}}
\def\Ib{{\bf{I}}}
\def\C{{\mathcal{C}}}
\def\qtild{{\tilde{q}}}
\def\Sch{\mbox{\tiny{SCH}}}
\def\pb{{\bf{p}}}
\def\pp{{\prime\prime}}
\def\c{s}
\def\s{i}
\def\qc{q}

\def\Gcal{\mathcal{G}}
\def\cb{{\bf c}}
\def\gw{\alpha}

\def\gwsm{\gw}

    \def\rr{q}
         \def\rri{d}
\def\em{e}
\def\ab{a}
\def\Ocal{\mathcal{O}}
\def\qz{q}
\def\pe{{\mbox{\scriptsize{+\hspace{-0.8mm}e}}}}
\def\nee{{\mbox{\scriptsize{-\hspace{-0.3mm}e}}}}
\def\qt{{qt}}
\def\fz{\phi}
\def\ft{\theta}

\def\Acalt{ {\widetilde{\mathcal{A}}} }
\def\Acal{\mathcal{A}}
\def\Ecal{\mathscr{E}}
\def\Tcal{\mathcal{T}}
\def\Acalw{\Acal'}
\def\A{A}
\def\vphiq{\psi_q}
\def\dash{{\mbox{\tiny{$_-$}}}}

\def\etam{\eta}

\def\b{\etam}



\def\bcal{{\hspace{-0.02cm}\eta\hspace{-0.17cm}_{\mbox{\tiny{$-$}}}\hspace{-0.05cm}}}
\def\bcals{{\hspace{-0.02cm}\eta\hspace{-0.17cm}_{\mbox{\tiny{$-$}}}\hspace{-0.03cm}}}
\def\bcalb{{\hspace{-0.02cm}\eta\hspace{-0.14cm}_{  \mbox{\tiny{$-$}}        }\hspace{-0.04cm}}}



\def\rhow{\rho'}
\def\engw{\eng'}
\def\engb{\tilde{\eng}_\vphi}
\def\engqa{\tilde{\eng}_q}
\def\engqan{\tilde{\eng}_{qn}}
\def\Aqa{\tilde{A}_q}
\def\enga{\tilde{\eng}}
\def\velqz{\vel_{qz}}
\def\rhoqz{\rho_{q}}
\def\Dqz{D_{q}}
\def\osub{{\mbox{\small{$0$}}}}
\def\N{{\hspace{-0.05cm}\mbox{\tiny{$N$}}}}
\def\Nsm{{\mbox{\footnotesize{$N$}}}}

\def\Ncal{{\cal N}}
\def\Ncr{\aleph}
\def\od{{\rm od}}
\def\eve{{\rm ev}}
\def\vtheta{\vartheta}
\def\phiz{\fz}
\def\vphiz{\phi}
\def\phix{\varphi}

\def\nh{n}
\def\fa{f}
\def\zrm{\mbox{\rm{z}}}
\def\minus{\mbox{-}}
\def\Rbqt{{\bf R}}
\def\a{\alpha}
\def\ph{{{\rm ph}}}
\def\ph{{\rm{ph}}}
\def\ub{{\bf u}}
\def\v{{\mbox{\scriptsize{\rm{v}}}\hspace{-0.03cm}}}
\def\vac{\v}
\def\strvac{\supvac}
\def\supvac{\v}
\def\Nsub{{\mbox{\tiny${N}$}}}
\def\Zsub{z}
\def\Xsub{x}
\def\Xssub{x}

\def\kin{{\rm kin}}
\def\Y{Z}
\def\Z{Z}
\def\Pcal{{\mathcal{P}}}
\def\Fcal{{b_m}}
\def\bfrak{b_\v}
\def\ii{n}
\def\Esub{{\mbox{\tiny${E}$}}}
\def\cEphi{\Ccal_{\Esub \vphi}}
\def\cEu{\Ccal}
\def\cuphi{A_q}
\def\Ccal{{\mathcal{C}}}

\def\Bcal{{\mathcal{B}}}
\def\Ucal{{\mathcal{U}}}
\def\subempty{\empty}
\def\str{\nus}
\def\strempt{{}}

\def\prim{\prime}
\def\ArticleLabel-lp{16}
\def\Authorname{J.X. Zheng-Johansson}
\def\Cyc{\zrm_\vphi}
\def\vir{{\rm vir}}
\def\AppA{}
\def\AppChDy{}
\def\AppB{}
\def\AppC{}
\def\citeUnif{xx}
\def\citeHDDMunich{5} 
\def\vel{\upsilon}
\def\velt{\vel}
\def\Dt{D}
\def\jt{j}

\def\Eb{{\bf{E}}}
\def\Bb{{\bf{B}}}
\def\obs{{\rm obs}}
\def\ev{\epsilon}
\def\ke{\kappa}
\def\Omegavel{\mathbin{{\mit\Omega}\mkern-13.mu^{_{\mbox{$-$}}}\hspace{-0.08cm}{}_d }}
\def\Wvel{\mathbin{{\mit\Omega}\mkern-13.mu^{_{\mbox{$-$}}}\hspace{-0.08cm}{}_d }}
\def\wvel{{\omega\hspace{-0.3cm}-\hspace{-0.02cm}}_d}

\def\q{\mathbin{q\mkern-11mu-}}
\def\empty{{\mbox{\tiny${\emptyset}$}}}
\def\nf{n}
\def\Kcal{{\math{K}}}
\def\Xb{{\bf{X}}}
\def\CL{{\mbox{\tiny{{\it C.L}}}}}
\def\imr{{\rm im}}
\def\pl{{\mbox{\tiny{$\|$}}}}
\def\Thm{\vartheta}
\def\TThm{{\mit{\Theta}}}
\def\Thetam{{\mit{\Theta}}}
\def\veli{{\mathscr{V}}}
\def\velib{{\pmb{\veli}}}
\def\Xima{\xi}
\def\Vel{W}
\def\Velb{{\bf{\Vel}}}
\def\Pw{{\psitot_p}}
\def\PW{{\Psim_p}}

\def\Mcal{{\mathfrak{M}}}
\def\Ncal{{\mathfrak{N}}}
\def\Ncalsub{\mbox{\tiny{${\mathfrak{N}}$}}}
\def\nablab{{\pmb{\nabla}}}
\def\velb{{\bf{v}}}
\def\rb{{\bf{r}}}
\def\jb{{\bf{j}}}
\def\fb{{\bf{f}}}
\def\Fb{{\bf{F}}}
\def\wrm{{\rm w}}
\def\med{{\med}}
\def\mfrak{{\mathfrak{M}}}
\def\Mfrak{{\mathfrak{M}}}
\def\Mfk{{\mathfrak{M}}}

\def\J{\mathcal{J}}
\def\Jcal{j}
\def\Jcalb{{\bf{j}}}
\def\Jobs{j_{\rm obs}}
\def\jobs{\Jobs}
\def\jr{j_{\rm obs}}
\def\obs{{\rm{obs}} }
\def\qm{{\rm qm}}
\def\Dscr{D_{\qm}}


\def\w{\omega{}}

        \def\Eng{ E^0{}}
        \def\eng{E}
        \def\engt{{\mathscr{E}}}

\def\engqi{\eng_{q}}
\def\engi{\eng}

\def\Qcal{\mathbin{{Q}\mkern-8.5mu^{_{\mbox{\small{$\dash$}}}}\hspace{-0.04cm} }}
\def\dash{{\mbox{\tiny{$_-$}}}}
\def\W{{\mit \Omega}}
\def\Eng{{\cal E}}

\def\engo{\bar{\eng}}
\def\Engo{{\bar{\Eng\hspace{0.1cm}}\hspace{-0.1cm}}}

\def\utot{\mathscr{U}}
\def\V{\mathscr{V}}
\def\Vt{\mathscr{V}}
\def\Vat{{\widetilde{V}}}
\def\Va{V}

\def\Vvq{V_{\v q}}
\def\Vvqo{V_{\v q0}}
\def\uscr{\mathscr{U}}
                \def\uscrz{\mathscr{U}}
                \def\uscrx{\mathscr{U}}
\def\uscrxb{{\pmb{\mbox{$\uscrx$}}}}

\def\uscrzb{{\pmb{\mbox{$\uscrz$}}\hspace{-0.08cm}}}

\def\uscrzt{z}
\def\uscrxt{x}

\def\uscrzo{z}
\def\uscrxo{x}

\def\zt{z}
\def\xt{x}
\def\Zt{Z}

\def\Xcal{\mathcal{X}}
\def\Zcal{\mathcal{Z}}
\def\uscrw{\uscr'}
\def\ubscr{\pmb{\mathscr{U}}}

\def\ubscr{\pmb{\mathscr{U}}}
\def\ubscrq{{\pmb{\mathscr{U}}\hspace{-0.14cm}}_q}
\def\ubscrqsq{{\pmb{\mathscr{U}}\hspace{-0.14cm}}_q{\hspace{-0.05cm}}}

\def\Zcal{{\mathcal{Z}}}
\def\Kcal{{\mathcal{K}}}
\def\kappab{\pmb{\kappa}}
\def\gd{{\mathcal{G}}}
\def\lb{{\bf l}}
\def\vb{{\bf v}}
\def\Rb{{\bf R}}
\def\pd{\partial}
\def\vphi{\varphi}
\def\psitot{{\mathcal{Y}}}
\def\psiR{\widetilde{\psi}}
\def\psiL{\widetilde{\psi}^{{\rm vir}}}
\def\psitotR{\widetilde{\psitot}}
\def\psitotL{\widetilde{\psitot}^{{\rm vir}}}
\def\PhimR{\widetilde{ {\mit \Phi}}}
\def\PsimR{\widetilde{ {\mit \Psi}}}
\def\PsimL{{\widetilde{ {\mit \Psi}}}^{{\rm vir}}}
\def\a{\alpha}
\def\apl{a}
\def\aplF{}
\def\uav{\bar{u}}
\def\D{\Delta}
\def\t{t}
\def\x{x}
\def\z{z}
\def\y{y}

\def\th{\theta}
\def\r{{\mbox{\tiny${R}$}}}
\def\re{{\mbox{\tiny${R}$}}}
\def\Fmed{F_{{\rm a.med}}}
\def\med{{\rm med}}
\def\Lw{L_{\varphi}}

\def\Ac{ \varphi}
\def\Tssub{{\mbox{\tiny${T}$}}}
\def\Kb{{\bf{K}}}
\def\kb{{\bf{k}}}
\def\Ksub{{\mbox{\tiny${K}$}}}
\def\W{{\mit \Omega}}
\def\Wd{\W_d{}}
\def\Nu{{\cal V}}
\def\Nud{\Nu_d{}}

\def\Acuni{\Ac_{{\Ksub}^\dagsup}^{\dagsup}}
\def\unduni{\Ac_{{\Ksub}^\dagger}^{\dagsup}}
\def\Acauni{\Ac_{{\Ksub}^\ddagsup}^{\ddagsup}}
\def\Acunim{{\Ac_{{\Ksub}^\dagsup}^{\dagsup *}}}
\def\undunim{{\Ac_{{\Ksub}^\dagsup}^{\dagsup}}^*}
\def\Acaunim{{\Ac_{{\Ksub}^\ddagsup}^{\ddagsup *}}}
\def\pd{\partial}
\def\Ad{ {\mit \psi}}
\def\psim{ {\mit \psi}}
\def\Kd{K_d{}}
\def\Lam{{\mit \Lambda}}
\def\lam{\lambda}
\def\dagsup{{\mbox{\tiny${\dagger}$}}}
\def\ddagsup{{\mbox{\tiny${\ddagger}$}}}
\def\psimKdK{\psim_{\Ksub,\Kdsub}}
\def\w{\omega{}}
\def\wdlow{\omega_d }
\def\g{\gamma{}} 
          \def\Xim{{\mit{\Xi}}}

\def\Phim{{\mit \Phi}}
\def\Psim{{\mit \Psi}}     
\def\Psima{{\mit\Psi}}

\def\arm{{\rm a}}
\def\brm{{\rm b}}
\def\crm{{\rm c}}
\def\drm{{\rm d}}
\def\erm{{\rm e}}
\def\frm{{\rm f}}
\def\grm{{\rm g}}
\def\hrm{{\rm h}}
\def\lf{\left}
\def\rt{\right}
\def\Kdsub{{\mbox{\tiny${K_d}$}}}
\def\psimkd{\psim_{\kdsub}}
\def\psimKd{\psim_{\Kdsub}}
\def\hquad{ \ \ } 
\def\Taum{{\mit \Gamma}}

\def\bbar{\hspace{-0.12cm}\mathbin{{b}\mkern-8.2mu^{{\mbox{\tiny{$-$}}}}\hspace{-0.11cm}}}
\def\cc{r}
\def\qq{\cc}
\def\oor{r}
\def\ho{}

\def\dagsup{{\mbox{\tiny${\dagger}$}}}
\def\ddagsup{{\mbox{\tiny${\ddagger}$}}}


\begin{abstract}
It is in accordance with such experiments as single photon self-interference that a photon, conveying one radiation energy quantum "$ h \times$ frequency", is spatially extensive and stretches an electromagnetic wave train. A wave train, hence an energy quantum, can only be emitted by its source gradually. In both the two processes the wave and "particle" attributes of the radiation field are simultaneously prominent, where an overall satisfactory theory has been lacking. This paper presents a first principles treatment, in a unified framework of the classical and quantum mechanics, of the latter process, the emission  of a single radiation quantum based on the dynamics of the radiation-emitting source, a charged oscillator which is itself extensive across its confining potential well. During the emission of one single radiation quantum, the extensive charged oscillator undergoes a continuous radiation damping and is non-stationary. This process is in this work treated using a quasi stationary approach, whereby the classical equation of motion, which directly facilitates the correspondence principle for a particle oscillator, and the quantum wave equation are established for each sufficiently brief time interval. As an inevitable consequence of the division of the total time for emitting one single quantum, a fractional Planck constant $h$ is introduced. The solutions to the two simultaneous equations  yield for the charged oscillator a continuously exponentially decaying Hamiltonian that is at the same time quantised with respect to the fractional-$h$ at any instant of time; and the radiation wave field emitted over time stretches a wave train of finite length. The total system of the source and radiation field  maintains at any time (integer $n$ times) one whole energy quantum, $h \times$ frequency, in complete accordance with the  notion of quantum mechanics and experiment.

\end{abstract}

\newpage

\section{Introduction}\label{sec-intr}

According to the classical electrodynamic theory (J C Maxwell, 1873), the  electromagnetic radiation fields are  manifestly waves. The theory  satisfactorily accounts for such processes as  diffraction,  interference and superposition. According to  quantum theory laid foundation to by M Planck in 1900,  electromagnetic radiation fields consist at the scale $h$ of energy quanta, and these are manifestly "particles" (M Planck, 1900; Einstein, 1905).  The theory is successful especially for such  processes where there presents an energy transfer between the radiation field  emitter and an external absorber, in general by one energy quantum (or photon),  $\hbar \w$ or by integral $n$ multiples of $n\hbar \w$ at a time. A photon, as well as a quantum matter particle,  is in the current interpretation of the quantum theory regarded as a statistical point particle. While the classical electrodynamics and the quantum electrodynamics  (Dirac, 1927) have proven extremely successful  where the wave and the particle attributes  are separately prominent, the two essentially parallel theories  present apparent clashes where the wave and "particle" attributes are simultaneously  prominent\cite{Bekefi Barrette,taylor1909}. A classical example of this is the single photon self-interference  in a double slit as demonstrated in experiments  \cite{taylor1909}. Here, the clash is a logical one:  a photon ($\hbar \w$), being regarded as a statistical point particle, logically can not pass two slits at the same time. A realistic theoretical representation of the (self) interference would require the radiation quantum to be depicted as a coherent (radial) wave train. 

A separate example is the intermediate process of emission (or absorption) of a radiation quantum $\hbar \w$, or photon; this is closely relevant to the single photon self-interference and will be the central concern of  this paper. 
Based on the classical electrodynamics as well as experiment, the radiation field is gradually and continuously emitted (or absorbed) by the source (or target) charge over a time duration $\sim 10^{-8}$ s \cite{Bekefi Barrette}, and therefore stretches a coherent wave train of a definite phase, a necessary pre-condition for producing single photon self interference. A requisite quantisation of the field however is not facilitated in this theory. According to the quantum theory and experiment, the radiation energy transfer from its emitting charge  to an external absorber is by one energy quantum (or less frequently by integral multiples of a quantum) at a time. The radiation energy quantum, when pictured as a statistical point particle and hence  a non-dividable unit, on the other hand,  can only be emitted (or absorbed) instantaneously (the observational finite elapsing time of a radiation decay instead is  attributed entirely to the statistical transitions of many sources, such as  atoms,   see e.g. H G Kuhn \cite{Bekefi Barrette}).  Similarly here, there presents a pronounced clash between the two theories. The clash, as we have seen from the two examples,  arises primarily  from the current interpretation of the quantum theory rested on an over simplifying statistical point particle picture, on top of an otherwise rigorous mathematical framework of the quantum mechanics. 

The wave and particle pictures may be readily reconciled with one another if we regard the radiation wave field (assuming an angular frequency $\w$) as being at any one time distributed across a finite distance, whence a wave train of a finite (non-zero) length and yet having one fixed quantum  of energy, $\hbar \w$, as a result of wave amplitude quantisation.  This description, in the case of matter wave, is entirely consistent with the  corresponding mathematical quantum wave function solution ($\Psim(x,t)$) to, for example, the Schr\"odinger  equation. Namely,   $\Psim(x,t)$ is extensive at any one time $t$ and the integration of energy density, being proportional to $|\Psim(x,t)|^2$, across (the space interval enclosing) the wave train leads to one (or integral $n$ times one) energy quantum. For the matter wave, an intrinsically extensive, internally electrodynamic (IED)  scheme has been recently proposed  by  the author (see e.g. a recent review in [\citejxzjied a]), in terms of which the primary outstanding difficulties,  including ones  associated with wave-particle duality, attributable mainly to  the statistical point particle picture in the current interpretation of the quantum theory, may be overcome. The IED  description for matter wave can be in principle translated to one for an (existing) photon, although indirectly and for stationary state only  in its current form, by imposing together the solutions to the Maxwell's equations for radiation field and to the Schr\"odinger equation for the radiation-emitting charge. This is mainly because when the continuous emission of one energy quantum $\hbar\w$ is in question, there requires a treatment which can facilitate both  (1) a non stationary process and (2) a differentiable $\hbar \w$. The two features are as yet not incorporated in the exiting IED representation. Currently, to the author's knowledge  a direct theory for the intermediate emission process of  single radiation quantum does not exist. The existing time-dependent perturbation quantum theory deals with  transitions between sharply defined  stationary levels without regarding the intermediate processes from which  the stationary levels are finally  reached.  

In this paper, we give a direct, first principles and relativistic treatment of the intermediate process of the emission of an electromagnetic radiation quantum in an unified framework of the classical and quantum mechanics. The treatment is instrumented by a quasi-stationary quasi-harmonic  approach to the radiation-emitting source, a charged oscillator, and the introduction of a fractional Planck constant $h$ which justification is to be provided through its capability of achieving a consistency both within the overall theory representation and between the theory and experiment. We elucidate in Sec. \ref{Sec-eom} in a unified framework  the classical and quantum equations of motion and solutions, in Sec. \ref{sec-cont-emi} the unified classical and quantum solutions underlining the continuous emission of a radiation quantum, and in Sec. \ref{Sec-trans} the transition probability. 
In the case of a charged matter particle oscillator, the relativistic radiation, in the sense of including  both the thermal and rest-mass energy radiations, is facilitated in terms of the IED particle model in Sec. \ref{Sec-5}.

\def\i{j}
\section{Equations of motion}\label{Sec-eom}


\refstepcounter{subsection}\label{Sec-CEDPM2} \label{Sec-2}
\paragraph{\ref{Sec-CEDPM2} Quasi harmonic motion}

We consider an extensive charged  object such as a quantum particle or liquid-like entity (see Sec. \ref{Sec-IED}),  of mass $\Mv$ and charge $q$, located about an equilibrium position $\Rb_0(X_{10},X_{20},X_{30} )$ in the three dimensional ($R^3$) vacuum. The object  was endowed with a mechanical energy in the past time by an external driving force which has ceased action before time $t=0$.  From time $t=0$ the object  is in spontaneous motion under  the action of an elastic restoring force $\Fb=-  \nablab V= -\beta \uscrxb$ along $X_j$-direction  ($\i=1,2,3$), where $V(\uscrxb)=\frac{1}{2}\beta \uscrxb^2$  and $ \uscrxb=(X_{\i_c} (t)-X_{\i_0})\hat{X}_\i$ is the  displacement  of  its mass centre $X_{\i_c}$ at time $t$ from  $X_{\i_0}$. In addition,  the object is acted on by a radiation damping force $\Fb_{r}= - \gw_\mus \Mv_\mus \frac{d \uscrxb_\mus}{d t}$ apparently attributable to a  viscous (elastic) vacuum medium;  $\gw_\mus (>0)$ is a damping factor or decay rate. 

The Hamiltonian of the system acted on by the force $\Fb_r$ will be time dependent. Supposing that  $\gw$ is  small  so that its Hamiltonian  density is in any brief time interval  $\D t$  constant, the extensive object  will behave in $\D t$ effectively  as if a rigid object  (whence facilitating  the correspondence principle in the case of a quantum particle). We may thus write down the Newtonian equation of motion for the rigid-like extensive object as a whole, $ \Mv \frac{d^2 \uscrxb}{dt^2}-(\Fb_{r} +\Fb_{{}})=0$.
Or, 
$$ \displaylines{
\refstepcounter{equation}\label{eq-Fqa}
\hfill  
 \frac{d ^2 \uscrz}{d \t^2} + \gwsm_\mus \frac{d \uscrz}{d t} + \w_\mus^2 \uscrz_\mus   =0, 
\quad \w_\mus^2=\frac{\beta_\mus}{\Mv_\mus}. 
\quad
                         \hfill (\AppA\ref{eq-Fqa})
\label{eq-iedmot1}
}$$
Equation (\AppA\ref{eq-Fqa}) has the general complex damped harmonic oscillation solution, 
$$\displaylines{
\refstepcounter{equation} \label{eq-uqApp}\label{eq-soly1b}  
\hfill
\uscrz^c  (t)
=\Acal_\mus  e^{-(\frac{\gw_\mus}{2}  +i\wp_\mus) t}, 
\hfill
 (\ref{eq-soly1b})
}$$
 where $\Acal_\mus$  is the oscillation amplitude at $t=0$; $\w_\mus^{\dagsup}=\sqrt{\w_\mus^2-\gw_\mus^2/4} $; $ \uscrz(t)={\rm Re}[ \uscrz_\mus^c ]=\mbox{ $\Acal_\mus$} e^{-\frac{\gw_\mus}{2}t }\cos(\wp_\mus \t)$ gives the physical displacement.  The initial phase is not relevant here and is in (\ref{eq-soly1b}) set to zero. Associated with the solution $\uscrx$,  the oscillator  $\mus$ has at any time $t$  a kinetic energy $\engt_{\mus kin}(t) =\frac{1}{2}\Mv_\mus ( \frac{d \uscrz_{\mus}(t)}{d t} )^2= \frac{1}{2}\Mv_{\mus } \w_\mus^2 \Acal_\mus^2 e^{-\gw_\mus t} 
(\frac{\wp_\mus}{\wo_\mus} \sin \wp_\mus t+
\frac{\gw_\mus}{2\wo_\mus} \cos \wp_\mus t 
)^2$, elastic potential energy $ V_\mus(\uscrx) =\frac{1}{2} \beta_\mus \uscrz^2_{\mus}(t)
 =\frac{1}{2}  \Mv_{\mus }\w^2_\mus \Acal_\mus^2e^{-\gw_\mus t} \cos^2 \wp_\mus t$ and total mechanical energy or Hamiltonian, assuming $\gw<< \w$ and thus $\w \dot{=}\w^\dagger$, 
$$\displaylines{
\refstepcounter{equation} \label{eq-engnx2pa}  \label{eq-engqb}
\hfill 
\engt_{\mus}(t) 
=\engt_{\mus kin}(t)+V_{\mus}(\uscrx) 
=
\eng e^{-\gw_\mus t}  \rho_{\mus 0}(t) 
\dot{=}\mbox{ $\frac{1}{2}$} \Mv_\mus \w_\mus^2 |\uscrz_\mus^c|^2, 
\quad \eng= \mbox{ $\frac{1}{2}$}  \Mv_{\mus } \w_\mus{}^2 \Acal_{\mus}^2,
 \hfill 
\cr 
\hfill
\rho_{\mus 0}(t)
=        
\left[(1+ \frac{\gwsm_\mus^ 2}{4\wo_\mus{}^2}) \cos^2 \wp_\mus t+(1- \frac{\gwsm_\mus^ 2}{4\wo_\mus{}^2}) \sin^2 \wp_\mus t + \gwsm_\mus \wp_\mus  \sin \wp_\mus t \cos \wp_\mus t\right] 
\dot{=}  |e^{-i \wo_\mus t}|^2
=1. 
\hfill 
 (\ref{eq-engqb})
}$$
Here, $|\uscrz^c|^2 \dot{=}\Acal_\mus^2 e^{-\gw_\mus t } |e^{-i \wo_\mus t}|^2 =\Acal_\mus^2 e^{-\gw_\mus t} $,  $|e^{-i \wo_\mus t}|^2 =\cos^2 \wo_\mus t+ \sin^2 \wo_\mus t=1$. The dynamical variables $\w$, $\Mv$, $\beta$, $\gw$, $\Acal$, $ \engt$, etc. in general all contain a relativistic effect, which evaluation will be illustrated for the  IED oscillator in Sec. \ref{Sec-IEDtot}. Identical energy solutions may be obtained from  solving Maxwell's equations for the radiation field emitted by charged object (\ref{app-damp}). 

Under the condition $\gw_\mus << \w_\mus$ which is well fulfilled in the radiation experiments of interest here,  about any time $t$  there  will in general exist a brief time interval $\D t_\mus$ satisfying  $\frac{2\pi}{ \w_\mus}  << \D t_\mus << \frac{2\pi}{ \gw_\mus}$, so that during $\D t_\mus$ the displacement (\ref{eq-soly1b})  effectively has a constant  amplitude  $\Acal_\mus e^{-\frac{1}{4}\gw_\mus      (t + t +\D t)}$ and hence is   {\it quasi harmonic}. Accordingly, as further combined with Eq. (\ref{eq-engnx2pa}a), the extensive oscillator  has in   $\D t_\mus$ effectively a constant Hamiltonian, and hence is {\it quasi stationary}, agreeing with our pre-condition for establishing Eq.  (\AppA\ref{eq-Fqa}).


\refstepcounter{subsection}\label{Sec-SHO-Quat} \label{Sec-3}

\paragraph{\ref{Sec-SHO-Quat} Quasi stationary flow motion}
Alternately, we may directly describe  the extensive oscillator of Sec. \ref{Sec-CEDPM2}   by a linear  probability density $\rho(x_\i,t)=|\psi(x_\i,t)|^2$ along the  $X_\i$ direction, where $\psi$ is a complex function for the  same reason as $\uscrx^c$  is complex and will serve as a natural  independent variable of  the Hamiltonian similarly as $\uscrx^c$ in  (\ref{eq-engnx2pa}a). The coordinate  $x_\i$ is related to $\uscrx$ as $x_\i  =X_\i -X_{\i_0}
=\D \uscrz_\mus+ \uscrz_\mus$, where $\D \uscrx_\mus=X_\i -X_{\i_c} $. 
With $\uscrx_\mus$ given by   (\ref{eq-soly1b}), and necessarily $ \D \uscrx =\D \Acal_\mus  e^{- \gw t}$  
for the oscillator being  assumed rigid-body like, we have 
$$\displaylines{
\refstepcounter{equation} \label{eq-xmu}
\hfill
x_\i= 
\Acal' e^{-\frac{\gw_\mus}{2} t} \cos({\w_\mus t}), \quad 
\Acal'= \mbox{$(\D \Acal_\mus+\Acal_\mus)$}. 
\hfill (\ref{eq-xmu})
}$$
The $\uscrx_\mus$ motion of the oscillator is associated with a  $\rho$- flow motion in the given $x_\i$-direction, with a flow velocity $\vel_\nus$ and flow rate 
$$\displaylines{\refstepcounter{equation} \label{eq-jq}
\hfill
\jt_\nus=\rho_\nus \velt_\nus 
=-\Dt_\nus[\psi_\nus^* (\nabla \psi_\nus) -(\nabla \psi_\nus^*) \psi_\nus ], 
\quad
\Dt_{\nus}=\frac{i {} \bcals_\nus}{b \Mv }
\hfill (\ref{eq-jq})
}$$
Here, $\Dt_{\nus}$  is an imaginary  diffusion constant (a general derivation is given in [\citejxzjied b]); $b=2$  for an oscillator whose $\uscr$ motion is described by the Newtonian  equation  (\AppA\ref{eq-Fqa}) as is with  $\psi_q, \psi_d$ of  Sec. \ref{Sec-5}.1, and $b=1$ by the Maxwell's equations as is with the $\psip$,  Sec. \ref{Sec-5}.2. $\bcals_\nus(t)$ is a new real variable to be determined.

Equation (\AppA\ref{eq-Fqa}) holds, in each brief time interval $\D t $ here, only if  $\rho$ satisfies in  $\D t $ the continuity equation
$$ \displaylines{\refstepcounter{equation} \label{eq-rhoq}\label{eq-contnu2a}
\hfill  \label{eq-contnu2b}
\frac{\pd \rho_\nus}{\pd t} + \nabla \jt_\nus - \Ocal_\nus \rho_\nus=0, 
\hfill (\ref{eq-contnu2b})
}$$
where $\Ocal_\nus=\frac{V_\nus(x_\i)}{i\bcalb_\nus}- \frac{V_\nus(x_\i)}{i\bcalb_\nus} 
$.
Substituting the  expressions for  $\rho_\mus$,  $j_\mus$, $D_\mus$ and $V$, Eq. (\AppB\ref{eq-rhoq}) is decomposed into two second order differential equations 
for $\psi_\mus $, $ \psi^*_\mus$, which for $\psi$ and $b=2$ is given as 
$$ \displaylines{
\refstepcounter{equation}\label{eq-engqnnew1}\label{eq-contnu3a}
\label{eq-contnu3-a}
\hfill  i\bcal_\mus \frac{\pd \psi_\mus}{\pd \t} =H_\mus \psi_\mus, 
\quad 
H_\mus
=  -\frac{\bcals_\mus^2 }{2\Mv} \nabla_\mus^2+ V(x_\i), \quad V(x_\i)= \frac{1}{2}\Mv  \w_\mus^2 x_\i^2;
 \hfill  (\ref{eq-contnu3-a})
}$$
the equation for  $\psip{} _\mus$ for the case of $b=1$ will be given by 
 (\ref{eq-eqtotwav}.b), Sec. \ref{Sec-5}.2.  It is easily seen that  the Hamiltonian $H$ in (\ref{eq-contnu3-a}) is separable into a term ($H_{\mus 0}$) associated with a potential $V_{\mus 0}(\xbar_\i)$ dependent  only on $\xbar_\i=  \Acal'{} \cos (\w t)$  and independent on radiation, and a term ($H_{\mus \Isub}$)  with a potential $V_{\mus \Isub}$ describing the source--radiation interaction, i.e.,  
$$ \displaylines{
\refstepcounter{equation} \label{eq-Xim2}
\hfill
  H_\mus=H_{\mus 0}+H_{\Isub \mus}, \quad 
H_{\mus 0}=  - \frac{\bcals_\mus^2\nabla^2 }{2\Mv{}_\mus} + V_{\mus 0}(\xbar_\i),   \quad
   V_{\mus 0} (\xbar_\i)=\frac{1}{2}  \Mv_\mus \w_\mus^2 \xbar_\i^2,
\hfill
\cr
\hfill 
H_{\mus \Isub}=V_{\mus \Isub}=V(x_\i)-V_0(\xbar_\i)= -V_0(\xbar_\i) (1-e^{-\gw_\mus t}). 
\hfill (\ref{eq-Xim2})
}$$

If up to the initial  time $t= 0 $ the oscillator $\mus$ has emitted no (net) radiation and we set for $t= 0 $ (as for $t<0$),
$$\displaylines{\refstepcounter{equation} \label{eq-Q}
\hfill
2\pi \bcals_\mus(0 ) = \b_\mus(0) =h, 
\hfill (\ref{eq-Q})
}$$
 Eq. (\ref{eq-contnu3-a}) identifies then with the usual Schr\"odinger equation, describing the quasi stationary harmonic oscillator during a brief time interval $\D t_\mus $ about  $t=0 $  here.  And the Eq. (\ref{eq-eqtotwav}), Sec. \ref{Sec-IED}, later describes the  total wave  equation[\citejxzjied c] for the IED particle system. Suppose that upon external perturbation the oscillator  begins at $t=0$ a (net) emission of  radiation and  is to emit one energy quantum $\hbar \w$. This is necessarily a gradual process based on experimental indications discussed in Sec. \ref{sec-intr}. Of the $\hbar \w$,  $\w$ is an intensive quantity and, when as the natural frequency of a specified mass and potential system mainly of interest in this paper,   is in general a fixed value ---if disregarding possible small broadening due to particle velocity variation, as known in theory and experiment. So  during the intermediate process of emitting one radiation quantum $\hbar\w$, $\bcals_\mus(t)$ is  inevitably  the time-dependent counterpart of $\hbar$.

For the extensive oscillator is  quasi stationary and hence its $H$ is  effectively constant in  $\D t_\mus$, $ \psi_\mus(x_\i,t)$ must be factorisable as  $ \psi_\mus(x_\i,t)=\phi_\mus(x_\i) \theta_\mus(t)$, 
 $
\theta_{\mus}= e^{-i \frac{\engtsub_{\mus}}{\bcal_\mus} t}$ and $\psi_\mus
=\phi_\mus  e^{-i \frac{\engtsub_{\mus}}{\bcal_\mus} t} $. 
Placing $\psi_\mus$  in  (\ref{eq-contnu3-a}),  we obtain $-i^2\bcal_\mus \frac{\engt_{\mus}}{\bcal_\mus} \phi_\mus \theta_{\mus}=H_\mus  \phi_\mus \theta_{\mus}$, or $\engt_{\mus} \phi_\mus =H_\mus  \phi_\mus$. The last equation may be rewritten as, with the substitutions of  $\xi_\i \equiv \sqrt{\frac{\Mv \wo_\mus }{\bcal_\mus } } \ x_\i $ and accordingly $\fz_\mus (\xi_\i) =e^{-\xi_\i^2/2}  \phix_\mus (\xi_\i)$,
$$ \displaylines{
\refstepcounter{equation}\label{eq-engqnnew}\label{eq-psiqy3p}
\hfill 
\nabla^2  \phix_\mus-2\xi_\i \nabla   \phix_\mus +2n_\mus  \phix_\mus=0, \hfill\cr
               \hfill \qquad 
\engt_\mus(t)  \rightarrow  \engt_{n_\mus}(t)=(n_\mus+\frac{1}{2})\bcal_\mus(t) \wo_\mus 
 \hfill (\AppB\ref{eq-engqnnew}) 
}$$
 Except that  $\bcal_\mus(t)$ is  in place of $\hbar$, Eq. (\AppB\ref{eq-engqnnew}a) formally is  identical to the time-independent Schr\"odinger equation for  harmonic oscillator and can thus be solved accordingly with respect to $\bcal(t)$.

\section{Continuous emission of a radiation quantum }
\label{sec-cont-emi} \label{Sec-4}

\def\musp{{\mu}}
\def\mus{{}}

$\engt_\mus$ of (\ref{eq-engnx2pa}a), and $\engt_{n_\mus}$ of (\AppB\ref{eq-engqnnew}b) after subtracting the $n_\mus=0$ term, $\eng_{0_\mus}$ (which does not radiate and hence not present in the Newtonian Eq. \ref{eq-engnx2pa}), are the same energy, hence  $\engt_\mus(t)=\engt_{n_\mus}(t)- \eng_{0_\mus} $. This is rewritten as, with (\ref{eq-engnx2pa}a) for $\engt_\mus$ and (\AppB\ref{eq-engqnnew}b)  for $\engt_{n_\mus}$, 
$$\displaylines{
\refstepcounter{equation} \label{eq-engnx2pa-p}
\hfill \qquad\quad
\engt_{ n_\mus}(t)- \eng_{0_\mus}
=\eng_{n_\mus}  \rho_{\mus 0}(t)e^{-\gw_\mus t}
\dot{=}
\eng_{n_\mus} \ai(t)
= n_\mus \bcal_\mus(t) \w_\mus \qquad (a)\hfill
\cr
\hfill \qquad\quad
\eng_{n_\mus}(=\engt_{n_\mus}(0))
=\mbox{ $\frac{1}{2}$}  \Mv{}_{\mus } \w_\mus{}^2\Acal_{n_\mus}^2 
+\eng_{ 0_\mus}=n_\mus\hbar \w_\mus+\eng_{0_\mus} \qquad (b) 
\hfill
\cr
\hfill \qquad\quad
 \ai(t)=e^{-\gw_\mus t}                   \qquad (c)
\hfill 
\cr
\hfill\qquad\quad
\Acal_{n_\mus} =\sqrt{n_\mus} \Acal_{1_\mus}, \quad  
\Acal_{1_\mus}=(\frac{2\hbar }{\Mv{}_{\mus }\w_\mus})^{1/2}            \qquad (d) \hfill
\cr
\hfill\qquad\quad
\bcal_\mus (t)
= \frac{\Mv{}_{\mus } \w_\mus \Acal_{n_\mus}^2}{2n_\mus} e^{-\gw_\mus  t}
= \hbar e^{-\gw_\mus  t} \qquad (e)
\hfill 
 (\ref{eq-engnx2pa-p})
}$$
where $ n_\mus=1,2,\ldots$.
 In (\ref{eq-engnx2pa-p}b), or (\ref{eq-engnx2pa}), $ \Mv{}_\mus$ and $\w_\mus$  are constants characteristic of the oscillator and potential  system; $\Acal_\mus$  only is subject to change upon  excitation and thus to quantisation, hence $ \Acal_\mus \rightarrow\Acal_{n_\mus}$. Eqs.  (\ref{eq-engnx2pa-p}d)  follow from the last two equations of (\ref{eq-engnx2pa-p}b), and Eq. (\ref{eq-engnx2pa-p}e) from the last two equations of (\ref{eq-engnx2pa-p}a), where $\frac{\Mv{}_\mus \w_\mus \Acal_{n_\mus}^2}{2n_\mus}=\hbar $ from  (\ref{eq-engnx2pa-p}d). For the (quasi harmonic) oscillator undergoing electromagnetic radiation here, $\eng_{0_\mus}=0$ based on comparison with the empirical Planck energy equation. (One might also  interpret this as the consequence  that the ground level energy $\eng_0$ is finite but is never emitted as electromagnetic radiation.)

With  (\ref{eq-engnx2pa-p}d) in (\ref{eq-xmu}) and (\ref{eq-soly1b}),  we obtain
 $x_\i \rightarrow x_{\i n} =\frac{\D \Acal_{n_\mus} +\Acal_{n_\mus}  }{ \Acal_{n_\mus} } \uscrx_{n} $ and 
$$ \displaylines{
\refstepcounter{equation}\label{eq-uscrxn}
\hfill  
\uscrx \rightarrow \uscrx_{n}
= \Acal_{n_\mus}
e^{-\frac{\gw_\mus}{2} t} \cos (\w_\mus t)
= (2\pi B_{n_\mus} \bcal_\mus)^{1/2} \cos (\w_\mus t), 
\hfill (\ref{eq-uscrxn})
}$$ 
where $B_{n_\mus}= \frac{n_\mus}{\pi \Mv{}_\mus \w_\mus} $. With (\ref{eq-engnx2pa-p}a) for $\engt_{n_\mus}(t)$, $x_\i \rightarrow$  $x_{\i n}$, $\xi_\i \rightarrow \xi_{\i n}=\sqrt{  \frac{\Mv{}_\mus\w_\mus}{\bcal_\mus}  } x_{\i n} $,  and the standard Hermit polynomial solution for $(\vphi_\mus \rightarrow) \vphi_{n_\mus}$ or the normalised $H_{n_\mus}$, we obtain the total eigen  function  solution for  (\AppB\ref{eq-engqnnew}a)
$$\displaylines{
\refstepcounter{equation} \label{eq-psiq1}
\hfill
\psi_{n_\mus}(\xi_{\i n_\mus},t)=\phi_{n_\mus}(\xi_{\i n_\mus}) \theta_{n_\mus}(t) 
=C_{n_\mus} H_{n_\mus} (\xi_{\i n_\mus})
e^{-\frac{1}{2}  \xi_{\i n_\mus}^2}e^{-in_\mus \wo_\mus t} 
\hfill (\ref{eq-psiq1})
}$$
Accordingly, $\rho_\mus(\xi_{\i n_\mus},t)=|\psi_{n_\mus}(\xi_{\i n_\mus},t)|^2=|\phi_{n_\mus}(\xi_{\i n_\mus})|^2$  which is independent of time, implying that indeed the extensive oscillator $\mus$  moves as a rigid-object in the way that its Hamiltonian density  $\engt_n |\psi_n(\xi_{\i_n},t)|^2$ during a brief time interval $\D t$ is everywhere constant in time, as presumed in  Sec. \ref{Sec-CEDPM2}. Supposing that $\psi_{n_\mus}(x_{\i n_\mus},t)$ is normalised in $[-\frac{L_\mus}{2}, \frac{L_\mus}{2}]$, where $L\sim \Acal$, there is then at any time $t$
$$\displaylines{
\refstepcounter{equation} \label{eq-psiq2}
\hfill
 \int^{\frac{L_\mus}{2} }_ {-\frac{L_\mus}{2} }  \engt_{n_\mus}(t)  |\psi_{n_\mus}(x_{\i n_\mus},t)|^2 d x_{\i n_\mus} 
= \engt_{n_\mus}(t) 
 \int^{\frac{L_\mus}{2} }_ {-\frac{L_\mus}{2} }   |\phi_{n_\mus}(x_{\i n_\mus} )|^2 d x_{\i n_\mus}
=  \engt_{n_\mus}(t)
\hfill (\ref{eq-psiq2})
}$$

The charged oscillator $\mus$  of a time dependent $\engt_n(t)$ begins according to  (\ref{eq-engnx2pa-p}a) at time $t=0$ a (net) radiation-emission, assuming $\gw_\mus >0$.  At a later time $t $ (assuming less than an equilibrium  time  $\teqmu$ which  may be in question for e.g. an oscillator enclosed between reflection walls), it  will have emitted a total amount of radiation energy given as, with $\eng_{r,n_{\mus}} \equiv \eng_{n_\mus}$, 
$$\displaylines{
\refstepcounter{equation} \label{eq-engnx2pp1-A}
\label{eq-engnx2pp}
\hfill  
 \engt_{r,n_{\mus }}(t)= \eng_{n_\mus} -   \engt_{n_\mus}(t)
= (1-e^{-\gw_\mus t })\eng_{n_\mus} 
=\aii (t) \eng_{r, n_{\mus}} 
=n_{\mus } \bcal_{r \mus}(t) \w_\mus \qquad (a) 
 \hfill 
\cr
\hfill
\aii(t)=\frac{\engt_{r,n_{\mus}}(t)}{\eng_{r,{n_\mus}}   } 
=1-\ai(t) \qquad (b)
\hfill 
\cr
\hfill\qquad
\bcal_{r \mus}(t)= \hbar (1- e^{-\gw_\mus t}) =\hbar -\bcal_\mus(t)\qquad (c)
\hfill (\ref{eq-engnx2pp1-A})
}$$ 
\begin{figure}[h]
\vspace{-0.1cm}
\begin{center}
\includegraphics[width=0.95\textwidth]{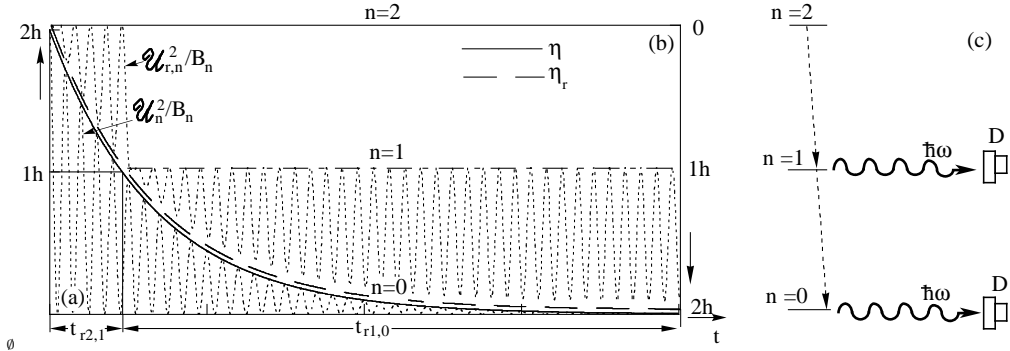}\end{center} 
\vspace{-0.7cm}
 \caption{ 
{\footnotesize
Continuous emission of a  radiation quantum $\hbar \w_\mus$ in a $(n_\mus=)2 \rightarrow 1 $ and $1 \rightarrow 0$ transition of a charged  oscillator. The intermediate processes are characterised by an exponentially  decreasing and increasing fractional Planck constants  of the  oscillator  and its radiation field as functions of time $t$, $\b_\mus(t)$ and  $\b_{r}(t)$  given by Eqs.  (\ref{eq-engnx2pa-p}e) and (\ref{eq-engnx2pp1-A}c), shown by the solid and long-dashed curves in graphs (a) and (b). The sinusoidal oscillation displacement squares  $\uscrz_{n_\mus}^2(t)/B_{n_\mus} $ and  $\uscrz_{r,n_\mus}^2(t)/B_{n_\mus} $   given after  Eq.  (\ref{eq-uscrxn}), short-dashed curves in (a) and (b),  are  modulated accordingly in amplitudes.
Used for the plot: $\w_\mus=2\pi/10$, $\gw_q=0.03$, $t_{r 2}\dot{=}2.3 \tau_\mus$, $t_{r 1} \dot{\ge} 200 -2.3\tau_\mus$. 
                %
         %
(c) is a schematic illustration of Fig \ref{fig-vibdecay.eps}a (dotted lines), and of Fig \ref{fig-vibdecay.eps}b (curly solid lines) as the permanently emitted electromagnetic wave trains of energy quanta $\hbar \w_q$'s, or  photons,  being probed by  detectors $D$.
}
}\label{fig-vibdecay.eps}
\end{figure}

It follows from Eqs. (\ref{eq-engnx2pa-p}) that during the intermediate process of emission of a radiation quantum, the $\engt_{n_\mus}(t)$, being  $\propto \uscrz_{n_\mus}^2(t)$ (dotted line  in Fig \ref{fig-vibdecay.eps}a), of the quasi-harmonic oscillator $\mus $ is at any instant of time quantised with respect to an exponentially decaying fractional-$h$, $\bcal_\mus(t)$ (solid line in Fig \ref{fig-vibdecay.eps}a). In this specific way, the oscillator emits electromagnetic radiation gradually and continuously with time. At the same time, as follows from Eqs. (\ref{eq-engnx2pp1-A}), the emitted radiation field $\psi_{r,n}$ has a Hamiltonian $\engt_{r,n_{\mus}}(t)$  that is quantised with respect to an exponentially  increasing  fractional-$h$, $\bcal_{r\mus}(t)$ (dashed line  in Fig \ref{fig-vibdecay.eps}b). The wave field $\psi_{r,n}$ emitted over time, being  propagated  in the vacuum medium at the speed of light $c$ outward from the source,  therefore stretches an extensive wave train, of a length $L'_{r\mus}(t)= c t$ at time $t$.   The sum of the two fractional-$h$'s of the source and radiation two-component system 
is at any time equal to 
$$\displaylines{
\refstepcounter{equation} \label{eq-bts}
\hfill
\b_\mus(t)+\b_{r \mus}(t)
=h e^{-\gw_\mus t} +h (1-e^{-\gw_\mus t})
\equiv h 
\hfill (\ref{eq-bts})
}$$ 
 (solid horizontal lines $n=0,1,2$ drawn through graphs  a and b in Fig \ref{fig-vibdecay.eps}). The total Hamiltonian is at any time $t$  given as 
$$ \displaylines{
\refstepcounter{equation}\label{eq-engtotmix}
\hfill  
\engt_{tot.n_\mus}(t)=\engt_{n_\mus}(t)+\engt_{r,n_{\mus}}(t) =\ai(t)\eng_{n_\mus} + \aii(t) \eng_{r,n_{\mus}}\equiv \eng_{n_\mus}\equiv \eng_{r,n_{\mus}}=n \hbar \w
\hfill (\ref{eq-engtotmix})
}$$ 
i.e. a constant consisting of $n$ quanta.

The energy difference ($\D \eng_{n, n-1}$) between adjacent stationary levels $n$ and $n'=n-1$ (given by Eq. \ref{eq-D-eng}b below)  is always one whole energy quantum $\hbar \w_\mus$, in complete accordance with the standard  notion of quantum mechanics and with overall experiments. This result may be also  stated as that, the difference action $\D \eng_{\mus n,n-1} \times \frac{2\pi}{\w_\mus}$ between two stationary levels is always equal to one full Planck constant $h$. This is shown in [\citejxzjied d] to be the combined consequence of the least action principle and second law of thermodynamics, i.e. the maximum entropy condition.

If the given  charged oscillator ($\mu$) is situated between (fixed) reflection walls, and no other absorbers  or perturbing fields present, the radiation wave field will be reflected back to $\mu$, be re-absorbed, and then re-emitted by it, iteratively. After an equilibrium  time $t = \teqmu$, the re-emission and re-absorption of radiation will reach equilibrium, the $ \engt_{r,n_{\mus }}(\teqmu), \ai(\teqmu), \b_{\mus}(\teqmu)$ etc. will be independent of time. We thus have a oscillator (as source) and radiation field two-component system    that is as a whole in stationary state and carry $n$ multiples of the one energy quantum $\hbar \w$, $n \hbar \w$, a situation as described by Eq. (\ref{eq-engtotmix}). The radiation field, being not charged nor undergoing energy transfer to external absorber (detector), is evidently not observable to an external observer (detector). The above in particular is the  scheme by which IED particle maintains as a distinct, stationary quantum system (Sec. \ref{Sec-IED}).

\section{Transition time and  probability}\label{Sec-trans}

Of the two-component system above,  the radiated fields will fail to be re-absorbed by their emission-oscillator $\mu$ if (i) no reflection walls present within distance of reach, (ii) another charged oscillator (as absorber) $\mu''$  nearby  begins to absorb the radiation emitted by $\mu$, and/or (iii) the oscillator $\mu$ is externally strongly disturbed away from stationary state. The radiation field will then become manifestly permanently emitted by $\mu$. 

Consider that the oscillator $\mu$ and radiation total system is at initial time $t=\timu$, after having previously undergone an equilibrium time  $\teqmu$, in the stationary  level $n_\mus$. Subjected to any of the circumstances (i)--(iii), the system begins from  time $t=\timu$ to undergo a permanent emission of a radiation quantum, transforming at  final time $t=t_{f\mus } = t_i +\taut_{n_\mus.n_\mus'} $ to the stationary level $n_\mus'$. Because in between $\teqmu$ and $t_i$ all the time dependent functions $\ai,\aii$ etc  remain the same, we may generally set   $t_i =\teqmu$; $t_i=\teqmu=0$ gives  an oscillator with no radiation at initial time. So at  $\tfmu=  \teqmu +\taut_{n_\mus.n_\mus'} $, the energy of the two-component system $\mu$ will have reduced by a total amount   given based on Eqs. (\ref{eq-engnx2pa-p}a,b), (\ref{eq-engnx2pp1-A}a) and (\ref{eq-engtotmix}) in two alternative ways as
$$\displaylines{
\refstepcounter{equation} \label{eq-D-enga}
\label{eq-D-eng}
\hfill\ 
\D \engt_{tot.n^{}_\mus,n_\mus'} 
= \engt_{tot.n_\mus}(\timu)- \engt_{tot.n_\mus}(\tfmu)
=\left[ \left(         e^{-\gw_\mus \teqmu } +\aii (\teqmu)\right)
- \left(e^{- \gw_\mus \tfmu} +\aii (\tfmu)\right) \right]\eng_{n_\mus} 
\hfill 
\cr    
\hfill\quad
=e^{-\gw_\mus \teqmu } (1- e^{- \gw_\mus \taumu })n \hbar \w_\mus 
=\D \engt_{n,n'}
\quad (a); 
\quad
\D \eng_{n_\mus.n_\mus'} =\eng_{n_\mus }-\eng_{n_\mus'}=(n-n')\hbar \w_\mus \quad (b)  
\hfill (\ref{eq-D-eng})
}$$ 
In going from the second to the third of  Eqs. (\ref{eq-D-enga}a) we assumed that during the transition any new radiation emitted by the source contributes to the permanently emitted radiation (i.e. $\D \engt_{n_\mus.n_\mus'} $) only and not to $\aii \eng_{\mus n}$, so $\aii (\teqmu)=\aii (t_f )= 1-e^{-\gw_\mus \teqmu}$. Further from the identity relation $\D \engt_{n_\mus.n_\mus'} = \D \eng_{n_\mus.n_\mus'} $   we obtain the transition time or lifetime evaluated from a finite $t_i(=t_0)$ and from $t_i=t_0=0$  respectively
$$\displaylines{
\refstepcounter{equation} \label{eq-tn1n}
\hfill 
 t_{n,n'}=t_f-\teqmu
=\frac{1}{\gw} \ln \frac{n e^{-\gw \teqmu}}{n'[1-\frac{n}{n'} (1-e^{-\gw \teqmu})]}
, \hfill
\cr
\hfill 
t_{n_\mus,n_\mus'    } =t_f-0
= \frac{1}{\gwsm_\mus} \ln \frac{n_\mus}{n_\mus'} \hfill (\ref{eq-tn1n})
}$$ 
 $t_{n_\mus,n'    } $, with $n'=n-1$ say,  informs the literary time span for the exchange of one photon  between two systems $\mu$ and $\mu''$  rather than  a statistical average, for  in accordance to the fundamental quantum principle that one energy quantum or  photon can either be exchanged between two systems as a whole, or not exchanged at all. For an oscillator endowed with $n$ energy quanta ($\eng_{n}=n\hbar \w$) at the beginning, the total  time $t_r$ required for transiting to final $n'=0$ level is 
$$\displaylines{
\refstepcounter{equation} \label{eq-tr-x1}
\hfill
t_r= \sum_{n'=1}^{n} t_{n', n'-1}=\frac{1}{f\gw}, 
\quad \frac{1}{f}= \sum_{n'=1}^{n} \ln\frac{n'}{n'-1} \hfill
(\ref{eq-tr-x1})
}$$ 
These have the  mean values  $\langle t_r\rangle = \frac{\int^\infty_0 t e^{-\gw t dt }     }{      \int^\infty_0 e^{-\gw t dt} } =\frac{1}{\gw}$ (the mean life time) and $\langle \frac{1}{f}\rangle  =1$. Since  for the systems considered $\gw << \w$, or $\frac{2\pi}{\gw}\sim t_{n_\mus,n-1}  >> \frac{2\pi }{\w}=\tau $, during a  transition time $t_{n,n-1}$  the oscillator in general continuously oscillates a large  $\frac{t_{n,n-1}}{\tau}>>1$ number of oscillation cycles.

Suppose there presenting a large $N$ number of identical oscillators  that at initial time all lie at energy level $n$  and will decay to final level $n'$ statistically   via $n\rightarrow n'$ transitions,  elapsing a transition time $t_{n,n'}$ given by  (\ref{eq-tn1n}) each, and emitting  a total $\Ncal_{n,n'}$ energy quanta or photons. The total apparent elapsing time is $T=\sum_{i=1}^N (t_{n,n'})_i =N t_{n,n'}$.  So $\Ncal_{n,n'}=\frac{T}{t_{n,n'}} =N$. The probability per unit  time that any oscillator makes a $n\rightarrow n'$ transition and  emits one photon is thus given as $$\displaylines{
\refstepcounter{equation} \label{eq-tn1n-rate}
\hfill 
 \gw_{n,n'} =-\frac{d \Ncal_{n,n'}}{\Ncal_{n,n'}d t_{n,n'}} =- \frac{1}{\Ncal_{n,n'}} \frac{T}{t_{n,n'}^2} = 
\frac{1}{t_{n,n'}}=\frac{\gw}{ \ln \frac{n}{n'}}
\hfill (\ref{eq-tn1n-rate})
}
$$
Specifically if $n_\mus=1 $ and $n_\mus' =n_\mus-1=0$,  Eqs.  (\ref{eq-tn1n}b) or similarly (\ref{eq-tn1n}a), and (\ref{eq-tn1n-rate}) give 
$$\displaylines{
\refstepcounter{equation} \label{eq-tn1n-p}
\hfill
t_{1,0}
=\infty, \quad  \gw_{1,0}=\frac{1}{t_{1,0}}=0;
\hfill (\ref{eq-tn1n-p})
}$$ 
and for $n=2$, $n'=1$,   $t_{2,1}=\frac{\ln 2 }{\gw} $, $\gw_{2,1}= \frac{\gw}{\ln 2}$; etc. For a fixed $\gw$, $t_{n_\mus,n-1} $ reduces, and $\gw_{n,n-1}$ increases with increasing $n$.

The transition probability may be more generally expressed in the usual terms of a source--radiation interaction Hamiltonian, our $H_\Isub$ given by Eq. (\ref{eq-Xim2}d) earlier. The total and the unperturbed Hamiltonians $H(=H_0+H_\Isub)$  and $H_0$ are given by Eqs. (\ref{eq-Xim2}a) and (b). 
The total (or ensemble [\citejxzjied e]) wave function at any time $t$ is of the general form $\psi_{en}(x_\i,t)=\sum_m b_m (t) \psi_m(x_\i)  e^{-i \frac{\engtsub_m(t)}{\bcal(t)} t} $, where $b_m(t)$ is the amplitude of state $m$; $\psi_{en}$ clearly  is also a solution to Eq.  (\ref{eq-contnu3-a}a). Substituting (\ref{eq-Xim2}a) for $H$  and the equation above for $\psi_{en} $  in the corresponding  equation of (\ref{eq-contnu3-a}a) we obtain
$$ \displaylines{
\refstepcounter{equation}\label{eq-tran-prob1}
        \sum_m (H_0+H_\Isub) b_m(t) \psi_m(x_\i)e^{-i \frac{\engtsup_m}{\bcal } t}   =
   i\bcal \sum_m \left[\dot{b}_m(t)   \psi_m (x_\i)e^{-i \frac{\engtsup_m }{\bcal } t}  - \frac{i\engt_m}{\bcal } b_m(t)   \psi_m(x_\i)e^{-i \frac{\engtsup_m}{\bcal } t} \right]
}$$
Subtracting $ \sum_m H_0 \psi_m(x_\i,t) =\sum_m \engt_m \psi_{m}(x_\i,t)$, multiplying $\psi_k^*(x_\i)$ from the left and integrating over all $x_\i$,  we obtain, for the eigen functions $\psi_m (x_\i)$'s being orthogonal,    
$$\displaylines{
\refstepcounter{equation}\label{eq-tran-prob2}
\hfill
          \sum_m b_m(t) \int \psi_k^*(x_\i) H_\Isub \psi_m(x_\i) d x_\i e^{-i \frac{\engtsup_m}{\bcal } t}  
 =  i\bcal \sum_m \dot{b}_m(t) \int \psi_k^*(x_\i)  \psi_m (x_\i) dx_\i e^{-i \frac{\engtsup_m }{\bcal } t}. 
 \hfill 
\cr
\hfill 
{\rm Or}\quad
\int b_n(t) \psi_k^*(x_\i) H_\Isub \psi_n(x_\i) d x_\i e^{-i \frac{\engtsup_n}{\bcal } t}  
=  i\bcal (t) \dot{b}_k(t) e^{-i \frac{\engtsup_k}{\bcal } t}, 
\hfill(\ref{eq-tran-prob2})
}$$
 assuming the oscillator to be in a definite energy state $n$ at  initial time $t_i$, and thus  $b_m=0$ for all $m\ne n$. Based on the solutions in Sec. \ref{Sec-4}, throughout  the intermediate process of  quasi stationary transition from  level $n$ to $k$ here across a  time duration $(t_i,t_f)$, there present one  component wave function which maintains precisely the same  as at initial time,  $\psi_n(x_\i,t)=\psi_n(x_\i) e^{-i \frac{\engtsup_n(t) }{\bcal(t) } t}=\psi_n(x_\i)  e^{-i \frac{\eng_n }{\hbar  } t} $; and the other (in a fashion as discussed after Eq. \ref{eq-D-eng}) as at the final time $t_f$, $\psi_k(x_\i,t)=\psi_k(x_\i) e^{-i \frac{\engtsup_k(t) }{\bcal(t) } t} =\psi_k(x_\i)  e^{-i \frac{\eng_k }{\hbar  } t} $. And the amplitudes of the two quasi stationary component states reduces and increases respectively with time as $\ai{}_{n}(t)$ and $ a_{r,n} \rightarrow a_k(t) =1-\ai{}_{n}(t)$. So with  $b_n(t)=\ai{}_{n}(t) =e^{-\gw t}$ (accordingly $b_k (t)=a_k(t)=1- b_n(t)$; i.e. we are here facilitated with the explicit time dependent functions $b_n(t), b_k(t)$ instead of  the usual perturbation approach),  and $\bcal =\hbar e^{-\gw t}$ of (\ref{eq-engnx2pa-p}e),  denoting $\w_{k n } =\frac{\engt_k(t)-\engt_n(t)}{\bcal(t)} =\frac{\eng_k -\eng_n}{\hbar}$,  (\ref{eq-tran-prob2}) is rewritten as 
$$\displaylines{
\refstepcounter{equation}\label{eq-tran-prob3}
\hfill  
\dot{b}_k = \frac{1}{i \hbar } H_{\Isub_{k n}} e^{ i \w_{kn} t}, \hfill\cr
\hfill 
H_{\Isub_{k n}}= \int  \psi_k^*(x_\i) H_\Isub  \psi_n(x_\i) d x_\i 
= -  V_{0_0 } (1-e^{-\gw t})\int  \left(\frac{x_{\i }}{\Acal_ne^{-\frac{1}{2}\gw t }}\right)^2   \psi_k^*(x_{\i } )  \psi_n(x_\i) d x_\i=H_{\Isub_0} \Xcal_{kn}, 
\hfill\cr
\hfill
\Xcal_{kn}= \int  \left(\frac{x_{\i }}{\Acal_n e^{-\frac{1}{2}\gw t }                 }\right)^2   \psi_k^*(x_{\i } )  \psi_n(x_\i) d x_\i,
\hfill (\ref{eq-tran-prob3})
}$$
where $V_{0_0}= \frac{1}{2}\Mcal \w^2 \Acal_n'{}^2$, $H_{\Isub_0} = V_{0_0} (1-e^{-\gw t} )$. Notice that  $H_{\Isub_0} $, the amplitude of source--radiation interaction Hamiltonian, or similarly $H_{\Isub}$, is gradually switched on from zero at initial time $t_i$ to maximum  $V_{0_0}$ at final time $t_f$;  and this is not necessarily a small or perturbation quantity. For an oscillator with an initial Hamiltonian  $\eng_{1}=2V_{0_0}  $ and a corresponding a final time  radiation-source interaction Hamiltonian $H_{\Isub_0}=V_{0_0} \cdot (1-0)$, for example, $H_{\Isub_0}$ is equal just  to  the (maximum)  potential or binding energy $V_{0_0}$ of the (quasi) Harmonic  oscillator.

The instantaneous transition probability per unit time in a brief time interval say $(t, t+\tau)$, with $\tau<< t_{n,k}$, in which $H_{\Isub_0}=V_{0_0} (1-e^{-\gw t})$ is essentially a constant, is given as
$$\displaylines{
\refstepcounter{equation}\label{eq-tran-prob4}
\hfill  
{\mathcal{P}_{n,k} } =\frac{1}{\tau} |b_k|^2, 
\quad b_k = \int^\tau_0 \dot{b}_k d t
= -\frac{H_{\Isub_{ kn}} (e^{i \w_{kn}\tau }-1)}{\hbar \w_{kn}} 
\hfill (\ref{eq-tran-prob4})
}$$
Suppose that $\w_{kn}$ may have a finite continuous dispersion. The probability per unit time integrated over the entire possible dispersion range ($-\infty, \infty$) of the transition energy  $\eng_{kn}=E_k-E_n= \hbar  \w_{kn}$, as evaluated at $\frac{1}{2} \tau$ at which $b_k =- \frac{H_{\Isub_{kn}} }{\hbar \w_{kn}} e^{i  \w_{kn} (\frac{1}{2}\tau)}  2i \sin (\w_{kn} \frac{1}{2}\tau ) $, is
$$\displaylines{
\refstepcounter{equation}\label{eq-tran-prob4}
\hfill
 \int^{+\infty}_{-\infty}  {\mathcal{P}_{n,k} } d \eng_{k n}  =
\frac{1}{\tau}  \int^{+\infty}_{-\infty} |b_k|^2 d \eng_{k n} 
=\frac{4}{\tau \hbar^2 }\int^{+\infty}_{-\infty} |H_{\Isub_{kn}}|^2 \frac{\sin^2 \frac{1}{2}\w_{kn}\tau }{\w^2_{kn}} d (\hbar \w_{kn})
\hfill
\cr
\qquad \qquad\qquad \qquad 
=\frac{2}{\hbar}|H_{\Isub_{kn}}|^2 \int^{+\infty}_{-\infty} \frac{\sin y^2}{y^2} d y
= \frac{2\pi}{\hbar}|H_{\Isub_{kn}}|^2
= \frac{2\pi}{\hbar}|H_{\Isub_{kn}}|^2
\hfill (\ref{eq-tran-prob4})
}$$
where $y=\frac{1}{2}\w_{kn} \tau$; $\frac{\sin y^2}{y^2} \propto   \delta(y)$, and $ \int^{+\infty}_{-\infty} \frac{\sin y^2}{y^2} d y=\pi$.   (\ref{eq-tran-prob4}) after dividing $\eng_{kn}$ out  is by definition equal to the $\a_{n,n'}$ earlier for $n'=k$. Except with $H_\Isub(t) $ being here an explicit function of time and describing the instantaneous source-radiation interaction at any time $t$ during the emission of a radiation quantum, the conclusions (\ref{eq-tran-prob3})--(\ref{eq-tran-prob4}) are formally as given based on the standard perturbation approach (see e.g. \cite{Schiff}).  The source-radiation interaction Hamiltonian at the final completion of emission of one quantum  is simply found  at  time $t_f$.

\section{IED particle oscillator}\label{Sec-IED}\label{Sec-5}


\refstepcounter{subsection}\label{Sec-5.1}
\paragraph{\ref{Sec-5.1} The kinetic and the total charge oscillations}    

\begin{figure}[hhtp]
\vspace{-0.4cm}
\begin{flushleft}
\hspace{-0.cm}
\includegraphics[width=0.6\textwidth]{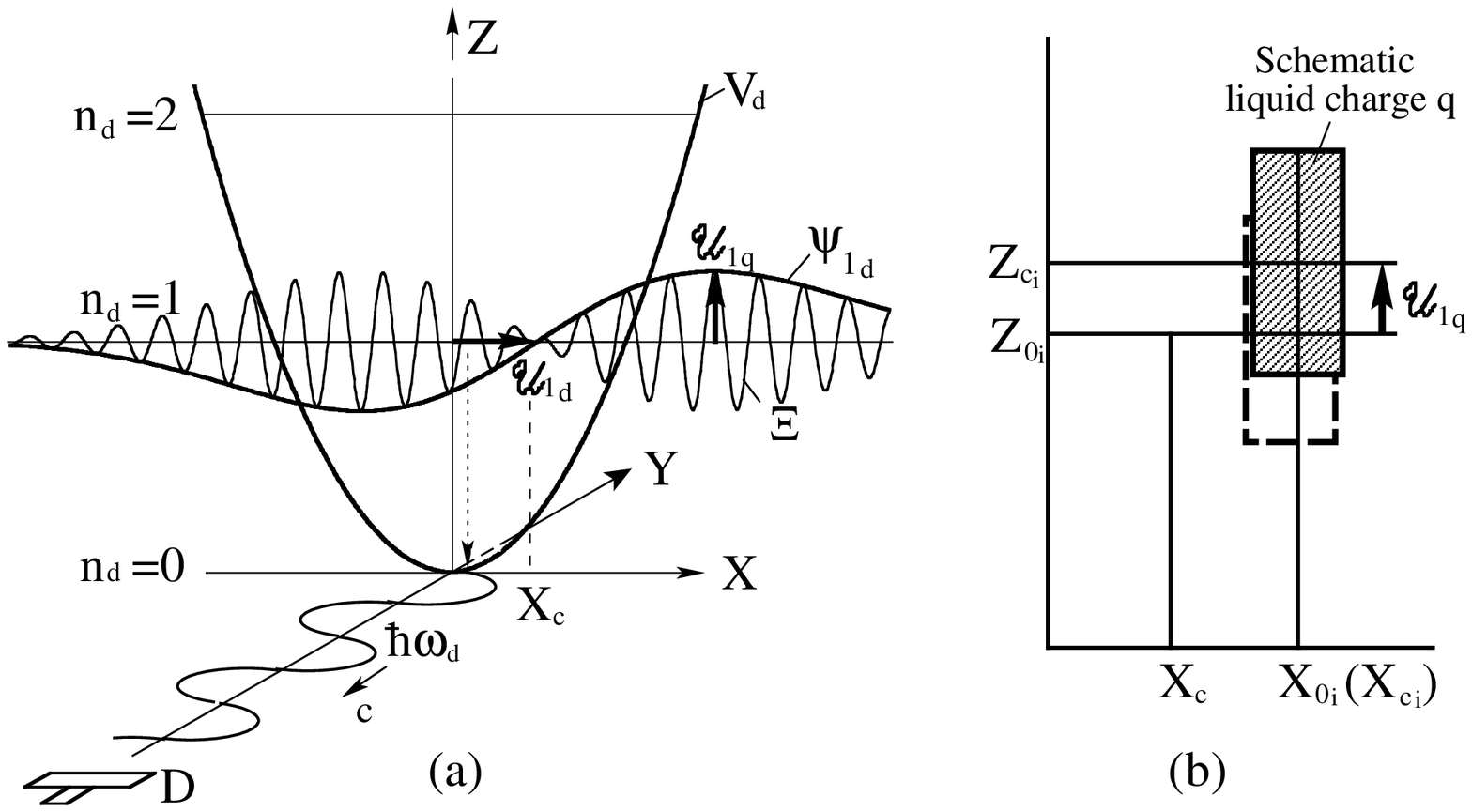} 
\end{flushleft}
\vspace{-5.2cm}
\begin{flushright}
\begin{minipage}[h]{4.cm}\baselineskip 0.35cm
{
\footnotesize
\refstepcounter{figure}\label{fig-sho-emit.eps}
Figure \ref{fig-sho-emit.eps}:  A quasi harmonic IED particle oscillator of oscillation $\uscrx_{n_d}$ (Figure \ref{fig-sho-emit.eps}a), which contains a  simultaneous internal oscillation $ \uscrx_{1_q}$ (Figure \ref{fig-sho-emit.eps}b), transits from initial thermal energy level $n_d=1$, with an eigen function  $\psi_{1_d}$ (solid curves in Figure \ref{fig-sho-emit.eps}a)  plotted after Eq. (\ref{eq-psiq1}), to a final level  $n_d=0$ ($\psi_{0_d}$ not shown).  At the end  one energy quantum $\hbar \w_d$ is emitted.
}\end{minipage}
\end{flushright} 
\vspace{ 0.cm}
\end{figure}

Consider as two inter-related applications  that  the (extensive quasi-harmonic) charged  oscillator of  Secs. \ref{Sec-2}--\ref{Sec-3} firstly represents an usual  charged oscillatory quantum particle of a charge $q$ and  mass $ m$. The particle has an oscillation  displacement $ \uscrx_d= X_{c }-X_0$ at its mass centre along the $X$ direction under the actions of an elastic force $ F_d =-\nabla V_d = -\beta_d \uscrx_d$ due to an applied potential  $V_d=\frac{1}{2}\beta_d \uscrx_d^2$    (Fig. \ref{fig-sho-emit.eps}a), and a radiation damping  force $ F_{rd}=-\a_d m \frac{d \uscrx_d}{d t} $.  Its kinetic motion is associated with a wave function $\psi_d(x,t)$,  where $x=X-X_0$. The equations of motion are  given directly by substitutions of $\uscrz_d$, $\gw_d$, $\w_d$, $\beta_d$, $ m$, $\psi_d$, $H_d$, $\bcal_d$ and $x$ for $\uscrz$, $\gw$, $\w$, $\beta$, $ \Mv$, $\psi$, $H$, $\bcal$ and $x_\i$ in Eqs. (\AppA\ref{eq-Fqa}) and  (\ref{eq-contnu3-a}),
$$ \displaylines{
\refstepcounter{equation} \label{eq-Fqa.a}
(\AppA\ref{eq-Fqa}.a)
\hfill 
 \frac{d ^2 \uscrx_d}{d \t^2} + \gwsm_d \frac{d \uscrx_d}{d t} + \w_d^2 \uscrz_d   =0; \quad  \w_d^2=\frac{\beta_d}{m}
 \hfill  (\ref{eq-Fqa.a})
\cr  \refstepcounter{equation} \label{eq-contnu3-a.a}
(\ref{eq-contnu3-a}.a) 
\hfill
 i\bcal_d \frac{\pd \psi_d}{\pd \t} =H_d \psi_d, 
\quad 
H_d=-\frac{\bcals_d^2 }{2m} \nabla_d^2  + \frac{1}{2}m  \w_d^2 x^2  
 \hfill  (\ref{eq-contnu3-a.a})
}$$
The solutions to  (\ref{eq-Fqa.a})--(\ref{eq-contnu3-a.a}) are given directly by substitutions  of the corresponding variables $n_d$, $\w_d$, $m$, $\Acal_{n_d}$, $a_d$, $\bcal_d$ and $\gw_d$ for $n$, $\w$, $\Mv$, $\Acal_n$, $a$, $\bcal$ and $ \gw $ in Eqs. (\ref{eq-soly1b})-- (\ref{eq-engqb}),  (\ref{eq-Q}) and (\AppB\ref{eq-engqnnew})--(\ref{eq-tr-x1}), Sec. \ref{Sec-4}. Of specific interest here, the Hamiltonian is given upon the substitutions above in Eqs.  (\ref{eq-engnx2pa-p}), divided for $t\le 0$ and $t\ge 0$ according to if any of the circumstances (i)--(iii) of Sec. \ref{Sec-trans} is onsets, as
$$\displaylines{
\refstepcounter{equation} \label{eq-engtotxd1}
\hfill
\hfill
  \engt_{n} (t)\rightarrow \engt_{n_d}(t) 
=\left\{
\begin{array}{c}
 \eng_{n_d}
= \frac{1}{2} m \w_d^2 \Acal_{n_d}^2  =n_d \hbar   \w_d  \qquad (t <0)   \cr
\engt_{n_d}(t)=a_d(t) \eng_{n_d}=n_d \bcal_d(t) \w_d \qquad (t\ge 0) 
\end{array}
\right. \hfill (\ref{eq-engtotxd1})
}$$
where $a_d(t)=e^{-\gw_d t}$, $\bcal_d(t)=\hbar e^{-\gw_d t}$. By (\ref{eq-engtotxd1}a), up to time $t=0$  the  particle is in accelerated ($\uscrx_{n_d}$)  motion but is in stationary state, with  a quantised Hamiltonian $\eng_{n_d}$. As such, the accelerated  (charged) particle does not radiate. The accelerated $\uscrx_{n_d}$ motion is instead to manifestly augment the frequency of the total wave from $\W_q$ by an average  factor $\g$ (given by Eq. \ref{eq-en-gtm}b below)  to $\w_q=\g \W_q$, and accordingly modulate the plane  electromagnetic wave $\Xim_{r,n_q}$   into  $\psi_{r,n_q}=\Xim_{r,n_q} \psi_{n_d}$ (Sec. \ref{Sec-5}.2). Under any of the circumstances (i)--(iii) of Sec. \ref{Sec-trans} only, the charged particle will  emit  thermal radiation according to (\ref{eq-engtotxd1}b).

We shall  be interested also in the rest-mass radiation of the particle. As a viable scheme for including the rest-mass radiation here, and for representing the extensive quantum wave of a matter particle in general, we shall represent the particle [assuming  for simplicity  (effectively) single charged and non-composite, like the electron] in terms of the IED particle model along with a vacuuonic vacuum proposed in  \cite{jxzjied} based on overall experiments. An IED particle  of charge $q$ is composed of a minute oscillatory  charge $q$ (as source) and the total electromagnetic radiation field $\Eb^j(\rb,t), \Bb^j(\rb,t)$ emitted by the charge. So,  besides the oscillation of the IED particle itself described by Eqs. (\ref{eq-Fqa.a})--(\ref{eq-contnu3-a.a}) in the $X$ direction here, there is simultaneously an internal oscillation of its generating charge $q$ (the $\uscrz_q$ along $Z$ direction below), and accordingly the dynamical process of its radiation field  (the $\psi_{r,q}$  in Sec. \ref{Sec-5}.2). The IED particle oscillation consists in the oscillation of the charge--radiation field   system as a whole (Sec. \ref{Sec-IEDtot}).

The vacuum is by construction (on experimental basis[\citejxzjied a,b]) filled of electrically neutral but polarisable vacuuons that are densely  and disorderly packed with a mean separation distance  $b_\v \sim 1\cdot 10^{-18}$ m. The vacuuons, polarised by the given charge $q$ situated in an interstice $i$ of the  vacuuons centred about $\Rb_{0_i}$, produce at the charge $q$ a vacuum potential  
$$\displaylines{
\refstepcounter{equation} \label{eq-Vvq}
\hfill 
              V_{\v q}(\uscrzb_q)=V_{\v q}(0) +\sum_n  \frac{1}{n!}\nabla^n \Vvq(\Rb_{0_i}) \uscrzb_q^n
\dot{=}V_{\v q0} + V_q, \quad 
V_{ q} 
= \frac{1}{2}\beta_q \uscrzb_q^2,  
\hfill
(\ref{eq-Vvq}) 
}$$
where $V_{\v q0}=V_{\v q}(0)$ and $\beta_q = \nabla^2 V_{\v q}$; $\uscrzb_q(t)=\Rb_{c_i}(t)-\Rb_{0_i}$ is the displacement  of the charge's  mass center $\Rb_{c_i}$ from $\Rb_{0_i}$ along a direction perpendicular to the maximum intensity of its radiation wave (the $\psi_{r,q}$ later, or the $\psi_d$ above along the $X$-direction); let this be the $Z$ axis  (Fig. \ref{fig-sho-emit.eps}.b); so $\uscrz_q(t) =Z_{c_i}(t)-Z_{0_i} $. The approximation in (\ref{eq-Vvq}a)  is given for $\uscrz_q$ being relatively small.

The charge  $q$ is a spinning liquid-like entity (or vortex) of a linear dimension $\sim b_\v \sim 1 \times 10^{-18}$ m; a point $Z$ on it has a displacement $z=Z-Z_{0_i}$, and probability density $\psi_q(z,t)$. The charge  has a zero rest mass. It however has a total mechanical energy (or Hamiltonian) $\engqi$ endowed in the past time, upon the action of  an external driving force which has ceased action before time $t=0$. The charge of the present time is thus an inertial system  moving about in the vacuum, spontaneously propelled by its own  inertial force $F_{ine} $. Defining  $F_{ine}\equiv \Mcal_q  \frac{d ^2 \uscrx_q}{d \t^2} $  in direct analogy to the Newtonian inertia for the usual matter-particles, we obtain $\Mcal_q$ as a proportionality constant, or  a manifestly dynamical mass upon mapping on to a non-viscous vacuum. Subjected to the vacuum  potential $V_{\v q}$ of Eq. (\ref{eq-Vvq}) about the fixed site $Z_{0_i}$ here, the motion of the charge  is resisted, by an elastic resistive force $F_{\v q}=-\frac{\pd V_{\v q}}{\pd \uscrz_q}=-\beta_q \uscrz_q$, and in addition a radiation damping force $F_{rq} = - \Mcal_q \a_q \frac{d \uscrz_q}{d t}$ in the viscous elastic vacuum. The corresponding equations of motion are  given directly by substitutions of  $\uscrz_q$, $\gw_q$, $\w_q$, $\beta_q$, $ \Mcal_q$, $\psi_q$, $H_q$, $\bcal_q$ and $z$ for $\uscrz$, $\gw$, $\w$, $\beta$, $ \Mv$, $\psi$, $H$, $\bcal$ and $x_\i$ in Eqs. (\AppA\ref{eq-Fqa}) and  (\ref{eq-contnu3-a}) as
$$ \displaylines{\refstepcounter{equation}\label{eq-Fqa.b}
(\AppA\ref{eq-Fqa}.b)\hfill 
 \frac{d ^2 \uscrx_q}{d \t^2} + \gwsm_q \frac{d \uscrx_q}{d t} + \w_q^2 \uscrz_q   =0, 
\quad \w_q^2=\frac{\beta_q}{\Mcal_q};
 \hfill (\ref{eq-Fqa.b})
\cr  \refstepcounter{equation}\label{eq-contnu3-a.b}
 (\ref{eq-contnu3-a}.b) \hfill
 i\bcal_q \frac{\pd \psi_q}{\pd \t} =H_q \psi_q, 
\quad 
H_q=-\frac{\bcals_q^2 }{2\Mcal_q} \nabla_z^2  + \frac{1}{2}\Mcal_q  \w_q^2 z^2  
 \hfill (\ref{eq-contnu3-a.b})
}$$
The Hamiltonian solution to Eqs. (\ref{eq-Fqa.b})-- (\ref{eq-contnu3-a.b}) is given similarly  by direct substitutions of $n_q$, $\w_q$, $\Mcal_q$, $\Acal_{n_q}$, $a_q$, $\bcal_q$ and $\gw_q$ for $n$, $\w$, $\Mv$, $\Acal_n$, $a$, $\bcal$ and $ \gw $ in Eqs.  (\ref{eq-engnx2pa-p}) as
$$\displaylines{
\refstepcounter{equation} \label{eq-engnq}
\hfill
\engt_n \rightarrow \engt_{n_q}
= \frac{1}{2} \Mcal_q \w_q^2 \Acal_{n_q}^2 a_{q}(t) =n_q \bcal_q(t)  \w_q,
\quad
\ai{}_q(t)=e^{-\gw_q t}, \quad \bcal_q(t)=\hbar e^{-\gw_q t} 
\hfill (\ref{eq-engnq})
}$$

In sum, for the IED particle oscillator, there present two simultaneous orthogonal (modes of)  oscillations $\uscrz=(\uscrx_d,\uscrz_q)$, and two corresponding probablity density $(|\psi_d|^2, |\psi_q|^2)$--flow motions in the $(X_\i,X_{\i'})=(X$,$ Z$) directions, executed by two apparent  oscillators, the IED particle oscillator $d$ itself  and the IED-particles's source charge $q$, of the masses $\Mv=(m, \Mcal_q)$ and a common charge $q$. The $d$ and $q$ oscillators each are  extensive in their respective potential wells $V=(V_d,V_q)$ of a common quadratic form and are described by the common forms of equations of motion, Eqs.   (\ref{eq-Fqa.a})--(\ref{eq-contnu3-a.a}) and  (\ref{eq-Fqa.b})--(\ref{eq-contnu3-a.b}). The solutions for the  two oscillators are accordingly formally commonly given by Eqs.  (\ref{eq-soly1b})-- (\ref{eq-engqb}),  (\ref{eq-Q}), and (\AppB\ref{eq-engqnnew})--(\ref{eq-tr-x1}).

\refstepcounter{subsection}\label{Sec-5.2}   \label{Sec-IEDWave}
\paragraph{\ref{Sec-IEDWave} The total wave motion}\label{Sec-IEDWave}
The charge oscillator $q$,  oscillating according to (\AppA\ref{eq-Fqa.b})--(\ref{eq-contnu3-a.b}) along the $Z$ direction in a three-dimensional vacuum, generates along any  radial  $\rb$-direction  two oppositely travelling,  and  $\vel_d$-motion resultant  Doppler-differentiated (Sec. \ref{Sec-IEDtot}) 
  electromagnetic wave fields $\Efb^j(\rb,t), \Bfb^j(\rb,t)$ travelling with velocities $\cb$ and $ -\cb$ parallel ($\j=\dagger$) and antiparallel ($\j=\ddagger$) to $\velb_d(t)$. Of direct relevance to the resultant IED particle is  the superposed total wave field   $\Efb (\rb,t)=\Efb^\dagger (\rb,t)+\Efb^\ddagger(-\rb,t)$; accordingly  $\Bfb=-\frac{\Ef}{c} \hat{\phi}$.
The $\Efb,\Bfb$ fields  in the case of  $V_d=0$ identify with  the solutions (\ref{eq-Era})--(\ref{eq-Erb}) to the Maxwell's equations,  \ref{app-damp}. In so far as the particle's coherent  wave motion,  and the associated total radiation power which is a constant  independent of  $r$ (Eq. \ref{eq-Pe}, \ref{app-damp}), are mainly in question here, the radial radiation waves $\Efb(\rb,t)$ may be in effect furthermore (i) represented by the maximum-intensity  wave field along the particle's motion $X$-direction and (ii) solved 
in  regions where $\rho_q=j_q=0$ only. For the  general case of a  finite $V_d$, wave fields  are of the general complex forms 
$$ \displaylines{
\refstepcounter{equation}\label{eq-Exc}
\hfill  
 \Efb_x^c(x,t) =\ai_q^{1/2}\Ef_{0_q} \psip(x,t) \hat{z}, \quad \Bfb_x^c(x,t)= -\frac{\Ef_{0_q} \psip(x,t) }{c} \hat{y}
\hfill (\ref{eq-Exc})
}$$ 
given similarly as in (\ref{eq-rhoeng-wave}), where  $\psip(x,t)=\psip^\dagger(x,t)+ \psip^\ddagger(x,t)$ is a complex dimensionless transverse wave and is to be solved in the presence of $V_d$.

The radiation field has a zero rest mass (similarly as its generating charge in the IED model) but a finite dynamical mass derived based on the following consideration. Suppose that at time $t$ the ratio of the radiation Hamiltonian ($\engt_{r,q}(t)$) to the total Hamiltonian of the charge-radiation system ($\engt_{tot,q}(t)$) is $a_{r,q}(t)$. Regardless of the applied $V_d$, the total radiated electromagnetic wave train (of the wave function $\psi_{r,q}$ and total length $L_{r,q}$),  propagating  at the constant velocity light $c$ in the  vacuum, has  according to Newtonian mechanics an (intrinsic)  total linear momentum $\la P_{r,q}\ra(=\frac{\int^{ \D  L_{r,q}}_0 \ev_0\Efb^2_{x}(x,t) d x}{c})= mc $ multiplied by $a_{r,q}^{1/2}(t)$, and kinetic energy $ a_{r,q}(t) \la \eng_{r,q,kin}(t)\ra (=a_{r,q}(t) \int^{\D L_{r,q}}_0\frac{1}{2}  \ev_0\Efb^2_{x}(x,t) d x) \ =a_{r,q}(t) \frac{ \la P_{r,q} \ra ^2 }{2m}=a_{r,q}(t) \frac{1}{2} m c^2$ (in  the electromagnetic energy expression in the brackets the factor $\frac{1}{2}$ in front of $ \ev_0\Efb^2_{x}$ is because  only the $\Efb$ field does work and not the $\Bfb$).   $m $ manifestly represents the relativistic dynamical mass of the wave train in a non viscous vacuum (as is assumed in Newtonian mechanics). In the vacuuonic vacuum representation we adopt in this section, $m$ represents a coefficient proportional to the resistive force $F_{\v r }$ of the viscous elastic vacuum bulk  against the wave train motion,
$\la P_{r,q}\ra =\int F_{\v r } d t =\int^c_0 m d c' $.
Accordingly, the wave train has in addition an elastic vacuum potential energy $a_{r,q} (t ) \la V_{r,q 0} (x)\ra= a_{r,q} \la \eng_{r,q,kin}(t)\ra  = a_{r,q}(t) \frac{1}{2} m c^2$, and therefore a total (intrinsic, i.e. excluding $a_{r,q} (t)  V_{d0}$) Hamiltonian 
$$\displaylines{
\refstepcounter{equation} \label{eq-mass-eng1}
\hfill \engt_{r,q}(t)- a_{r,q} \la V_{d0}\ra =  a_{r,q} \la \engt_{r,qkin}(t) \ra +a_{r,q}  \la V_{\v r,q0} (x)\ra  
=a_{r,q}(t)  2\times \frac{1}{2} m c^2 =a_{r,q}(t) mc^2 \hfill\cr
\hfill  
{\rm Or} \quad 
\hbar \w_q -\la V_{d0}\ra = m c^2 \hfill (\ref{eq-mass-eng1})
}$$
given after substituting the Eq. (\ref{eq-engrnq}) below for $\engt_{r,q}(t)$ and  dividing  $a_{r,q}(t)$ out. In the formal sense of the Eq. (\ref{eq-engtotx1}) later, $m$ represents the dynamic mass of the IED particle. 

The Maxwell's equations for the fields (\ref{eq-Exc}) lead in regions where $\rho_q=j_q=0$ to the wave equation  $ \frac{\pd^2 \psip}{\pd t^2}=(c^2 +\frac{V_d}{m}) \nabla^2 \psip$. This further reduces to, by combining with the identity relation  $m =\frac{\hbar \w_q -V_d}{c^2}= \g M$ described by Eqs. (\ref{eq-mass-eng1}b) (or  \ref{eq-engtotx1}) and (\ref{eq-en-gtm}) and   with  the procedure described in [\citejxzjied c]  except with $\bcal_{r,q}$ replacing  $\hbar$ in the final result,
$$\displaylines{
\refstepcounter{equation} \label{eq-eqtotwav}
\hfill
 i\bcal_{r,q} \frac{\pd \psip_\mus}{\pd \t} =H_p \psip_\mus, 
\quad 
H_p=-\frac{\bcals_{r,q}{}^2 }{m} \nabla_\mus^2
          +V_d
\hfill (\ref{eq-eqtotwav})
}$$
It is easily  seen that  Eq. (\ref{eq-eqtotwav}) is associated with a continuity equation, $\frac{\pd \rho_{r,q}}{\pd t} +\nabla (-D_p \nabla \rho_{r,q}) -(\frac{V_d}{i \bcalb_{r,q}}-\frac{V_d}{i \bcalb_{r,q}}) \rho_{r,q} =0$, given by substitutions of $\rho_{r,q}=|\psip(x,t)|^2$,  $b=1$, $m$, $D_p=\frac{i\bcalb_{r,q} }{m }$,  $V_d$ and  $x_\i$  for $\rho$, $b$,  $\Mv$, $D$, $V$ and $x$  in (\ref{eq-contnu2b}).

Using  the procedure of [\citejxzjied c] except with  $h$ being here replaced by the $\bcal_{r,q}$,  with $\psip = \Xim_{r,q} \psi_d$, Eq.  (\ref{eq-eqtotwav}) may be decomposed into two separate wave equations  for $\psi_d $ and $\Xim_{r,q}$.  The wave equation for $\psi_d(x,t)$,   with a Hamiltonian  $H_d=H_p-H_{p }^0 = - \frac{\bcals_{r,q}^2}{2m}  \frac{\pd^2 }{\pd x^2} + V_d$,  is just the Eq.  (\ref{eq-contnu3-a.a}). $\Xim_{r,q}$ is the total radiation field  emitted by  the charge $q$ when oscillating about a fixed site, i.e. $\vel_d(t)=0$, and is described by the wave equation 
$$ \displaylines{
\refstepcounter{equation}\label{eq-contnu3a-p}
\label{eq-engqnnew1}
\hfill   i\bcal_{r,q} \frac{\pd \Xim_{r,q} }{\pd t } = H_{p}^0 \Xim_{r,q}, 
\ \ H_{p}^0 = - \frac{\bcals_{r,q}^2}{M} \frac{\pd^2}{\pd x^2}.  
\hfill (\ref{eq-contnu3a-p}) 
}$$
(\ref{eq-contnu3a-p}) has  the solutions
$$\displaylines{
\refstepcounter{equation} \label{eq-Xim}
\hfill 
\Xim_{r,n_q}(x,t)=  e^{i(\frac{\Ptsub^0_{r,n_q}}{\bcal_{r,q}} x-\frac{\engtsub^0_{r,n_q}}{\bcal_{r,q}} t)}, 
\quad 
 \engt_{r,n_q}^0  =\lim_{\vel_d=0} \engt_{r,n_q}(t), 
\quad \Pt_{r,n_q}^0=\frac{\engt^0_{r,n_q}}{c},  
\hfill (\ref{eq-Xim})
}$$ 
The total Hamiltonian of the radiation field  is given after (\ref{eq-engnx2pp1-A}) as 
$$\displaylines{
\refstepcounter{equation} \label{eq-engrnq}
\hfill
\engt_{r, n}(t) \rightarrow \engt_{r, n_q}(t)
=\aii{}_{,q}(t) n_q \hbar \w_q =n_q \bcal_{r,q}(t)\w_q,
\hfill
\cr \hfill 
\aii{}_{,q} (t)= 1-\ai{}_q(t)= 1-e^{-\gw_q t}, \quad \bcal_{r,q}(t)=\hbar - \bcal_q(t)= \hbar (1-e^{-\gw_q t})
\hfill (\ref{eq-engrnq})
}$$


\refstepcounter{subsection}\label{Sec-IEDtot}
\paragraph{\ref{Sec-IEDtot} The charge and radiation-wave total system}

The  minute liquid-like charge $q$ and the  resulting radiation wave $\psip$ 
 are maintained as  one system, the IED particle, by the repeated radiation re-absorption and re-emission scheme commented after Eq. (\ref{eq-bts}). The total Hamiltonian  $\engt_{tot.n_q}(t)$ of the IED particle thus is at any time $t$ carried a fraction $a_{q}(t)$ by the charge oscillator $q$, and  $a_{r,q}(t)$ by the total radiation field $\psip$. With the $ \engt_{n_q}$ of  the charge and $\engt_{r, n_q}$ of  its radiation field given by (\ref{eq-engnq}) and  (\ref{eq-engrnq}),  we obtain, for $n_q=1$,
$$\displaylines{
\refstepcounter{equation} \label{eq-engtotx1}
\hfill
\engt_{tot.1_q}(t)=\engt_{1_q}(t)+\engt_{r,1_q}(t) =a_{q}(t)\eng_{1_q} + a_{r,q\mus}(t) \eng_{r,1_q}
 \hfill
\cr
\hfill 
\equiv \eng_{1_q}\equiv \eng_{r,1_q}
= \frac{1}{2} \Mcal_q \w_q^2 \Acal_{1_q}^2
= \hbar \w_q =mc^2 +V_{d0}. 
\hfill (\ref{eq-engtotx1})
}$$
The one energy quantum $ \hbar \w_q $ of the $q$--$\psip$ system will, applying  the general  quantum mechanical principle once again, either  not be absorbed by an external source at all, or be absorbed as a whole upon an  energy exchange.

As elucidated in [\citejxzjied b], an oscillatory charge $q$ (of a zero rest mass) at the stationary   level $n_q=1$ in  $V_{\v q }$  gives rise to a stationary  electron ($e$) if $q=-e$, and a proton ($p$) if $q=+e$. A transition of the $-e$ or $+e$ oscillator from initial level $n_q=1$ to a final $n_q'=0$  (in $V_{\v q }$) in the vacuum  corresponds to a spontaneous decay of the particle $e$ or $p$, over an infinite transition time $t_{1,0}=\infty$ according to Eqs. (\ref{eq-tn1n})--(\ref{eq-tn1n-p}), in direct  accordance with the empirical fact that the proton and electron  have under normal conditions infinite lifetimes. (Nevertheless, the $e$ or $p$  may transit to  the ground state $n_q=0$  if the quadratic $V_{\v q}$ condition is strongly distorted, as e.g. would be the case when an anti-particle presents nearby, leading to  a pair annihilation.) 
In a vacuum composed of densely packed vacuuons,  the $V_{\v q}$ of each site extends only about half way to its neighbouring site and is whereof superseded by $V_{\v q}$  of the neighbouring site. So there is no stationary state for $n_q>1$. The charge of an  existing electron,  proton or an composite particle of these,  carries already  one quantum of the  relativistic mass energy, $\hbar \w_q$, of the particle and can not absorb another  $\hbar \w_q$. So, the charge of another such existing  particle   can  only permanently emit its $\eng_{tot. 1_q}$, in a quasi stationary process, through  pair annihilation; and the emitted gamma photon  can only be absorbed by another "bare" charge (a vaculeon composing a vacuuon) out of the vacuum.

If any of the circumstances (i)--(iii) of Sec. \ref{Sec-trans} sets in (at time $t=0$)  so that the IED particle is perturbed away from  the stationary  level $n_d$, the IED particle oscillator  will now (tending to stabilise in a lower and stationary level) manifestly emit thermal radiation ($\psi_{r,n_d} $) according to Maxwellian electrodynamics, or equivalently the quasi harmonic solution Eq. (\ref{eq-engtotxd1}b). The emitted thermal radiation  energy is accordingly given as $\engt_{r,n_d}(t)  =\eng_{n_d}-\engt_{n_d}(t)= \eng_{n_d}(1-e^{-\gw_d t})=\bcal_{r,d} \w_d$, where $\bcal_{r,d}(t)= \hbar -\bcal_d(t)$  given after  (\ref{eq-engnx2pp1-A}) and  (\ref{eq-engtotxd1}).  The maximum intensity,  of the thermal radiation is in a direction perpendicular to $\uscrx_{n_d}$, hence lying  along a line in the $Y-Z$ plane, passing $(X_0,Z_0)$    in Fig \ref{fig-sho-emit.eps}. 

The sum of the two thermal terms  above,
$
\engt_{tot.n_d}(t) 
=\engt_{n_d} (t)+\engt_{r,n_d}(t)
= \eng_{n_d} e^{-\gw_d t}+ \eng_{n_d}(1-e^{-\gw_d t})
$, 
 represents two briefly co-existing  components during the transition only, rather than a distinct system given in (\ref{eq-engtotx1}), for the thermal frequency  $\w_d$ is not unique but is one out of a continuous spectrum  and the energy quantum $\hbar \w_d$ can be readily absorbed/emitted by another particle oscillator in the surrounding. At the end of the transition time $t_{n_d,n_d-1}$, the energy exchanged is given based on (\ref{eq-D-eng}) as $ \D \engt_{n_d,n_d-1}=\eng_{n_d} (1-e^{-\gw_d t_{n_d,n_d-1}}) = \D \eng_{n_d,n_d-1}=\hbar \w_d   $, i.e. one whole energy quantum. For the thermal electromagnetic  radiation here, the result $t_{\qq.1,0} =\infty$ of Eqs. (\ref{eq-tn1n})--(\ref{eq-tn1n-p}) implies that the IED particle oscillator will always maintain at least a "zero point" (ground-state) thermal energy, $\eng_{0_d}$.

We finally evaluate the relativistic effect due to the  $\vel_{n_d}$ (or $\uscrx_{n_d}$) motion of the particle in $n_d$th thermal level combining with the thermal and total energy solutions already obtained. The stationary  harmonic oscillation (given for $\gw_d=0$) of the (IED) particle is an accelerated  motion. The instantaneous velocity  $ \vel_{n_d}(t) $, accordingly the instantaneous relativistic mass $m(t)$,  kinetic energy and linear momentum
$$\displaylines{
\refstepcounter{equation} \label{eq-eng-kind}
\hfill
\engt_{kin, n_d}(t)=\frac{p_{n_d}^2(t)}{2 m(t)}=\frac{1}{2}m(t) \w_d^2 \Acal_{n_d}^2 \sin^2 (\w_d t ), \quad p_{n_d}(t) =m(t) \vel_{n_d}(t),
\hfill (\ref{eq-eng-kind})
}$$
and  potential energy $V_{n_d}(t)$ (as a short-hand denotation of  $V_{n_d}(x(t))$ here) as measured in the  laboratory frame ($S$), each  vary with time. In a brief time interval about $t$, $V_{n_d}(t)$ is effectively constant. Thus the relationship between the instantaneous total (internal) Hamiltonian $ \engt_{tot. 1_q}- V_{n_d}(t)$ (with $\engt_{tot. 1_q}(t) \equiv \eng_{tot. 1_q}=\hbar \w_q$ as given in Eq.  (\ref{eq-engtotx1}))  and the  $p_{n_d} (t)$  may be written  according  to the solution for a  IED particle of constant potential  (see e.g. [\citejxzjied c]), or equivalently the Einstein mass energy relation, as
$$\displaylines{
\refstepcounter{equation} \label{eq-eng-totx1}
\hfill
\left[\hbar \w_q(t)-V_{n_d}(t)\right]^2 = m^2(t)c^4= M^2c^4 + p_{n_d}^2(t) c^2,
\hfill (\ref{eq-eng-totx1})
\cr
\mbox{where  }
 \hfill
\cr
\refstepcounter{equation} \label{eq-eng-totx2a}
\hfill
\w_q(t)=\g(t) \W_q, \quad
 m(t)=\g(t) M, \quad 
V_{n_d}(t)=\g(t)V_{n_d}^0(t), \hfill
\cr 
\hfill
p_{n_d}(t)=\g(t) p_{n_d}^0(t)=\g(t)M\vel_{n_d}(t),
 \hfill       (\ref{eq-eng-totx2a})
\cr
\refstepcounter{equation} \label{eq-eng-totx2}
\hfill
\g(t)=\frac{1}{\sqrt{1-\vel_{n_d}^2(t)/c^2}}, \quad 
\g^2(t)=1+ \g^2(t)\frac{\vel_{n_d}^2(t)}{c^2}; 
\hfill (\ref{eq-eng-totx2})
}$$
 $\W_q$, $M$,  $V_{n_d}^0(t)$ and $p_{n_d}^0(t)$ are the corresponding 
rest-mass values measured in $S$.

The time average of  (\ref{eq-eng-totx1}) is $\la\left[\hbar \w_q(t) -V_{n_d} (t)\right]^2\ra = \la m^2(t)\ra c^4= M^2c^4 + \la p^2_{n_d}(t)\ra c^2$. The first two expressions develop as $\la\left[\hbar \w_q(t) -V_{n_d} (t)\right]^2\ra  =\la \g^2(t)\ra \left[\hbar \W_q-V^0_{n_d}  \right]^2 $ and $\la m^2(t)\ra =\la \g^2(t)\ra M^2$, where 
$$\displaylines{
\refstepcounter{equation} \label{eq-en-gt2p}
\hfill
\la\g^2(t)\ra= \left\la\frac{1}{1-\vel_{n_d}^2(t)/c^2} \right\ra=1+ \frac{\la \vel_{n_d}^2(t)\ra }{c^2}
+ \frac{\la \vel_{n_d}^4(t)\ra }{c^4} + \ldots
\hfill (\ref{eq-en-gt2p})
}$$
The $\la \vel_{n_d}^2(t)\ra$, $\la \vel_{n_d}^4(t)\ra$, etc. may be readily individually evaluated as 
$$\displaylines{
\refstepcounter{equation} \label{eq-eng-q-veld1}
\hfill
{\langle \vel_{n_d}^2(t) \rangle}
= \frac{\frac{1}{2}   m(t) \w_d^2 \Acal_{n_d}^2 \frac{1}{\tau_d}  \int^{\tau_d}_0 \sin^2 \w_d t d t            }{\frac{1}{2}  m(t)}
= \frac{1}{2}  \w_d^2  \Acal_{n_d}^2,
\hfill (\ref{eq-eng-q-veld1})
}$$
and ${\langle \vel_{n_d}^4(t) \rangle}=\frac{1 \cdot 3}{2\cdot 4}  \w_d^4 \Acal^4_{n_d}$, etc., where $\int^{\tau_d}_0 \sin^2 \w_d t d t =\frac{\tau_d}{2}$, $\int^{\tau_d}_0 \sin^4 \w_d t d t =\frac{1 \cdot 3}{2\cdot 4} \tau_d$, etc. Multiplying (\ref{eq-en-gt2p}) by $ \frac{\la \vel^2_{n_d}(t)\ra }{c^2}$  and in turn  adding 1 on each side, we obtain $\la\g^2(t)\ra \frac{\la \vel^2_{n_d}(t)\ra }{c^2} +1 =1+ \frac{\la \vel_{n_d}^2(t)\ra }{c^2} \frac{\la \vel^2_{n_d}(t)\ra }{c^2}
+ \frac{\la \vel_{n_d}^4(t)\ra }{c^4} \frac{\la \vel^2_{n_d}(t)\ra }{c^2} + \ldots$, or 
$$\displaylines{
\refstepcounter{equation} \label{eq-en-gt3}
\hfill
\la \g^2(t) \ra = 1+ \la \g^2(t) \ra  \frac{\la \vel^2_{n_d}(t)\ra }{c^2} = 1+ \la \g^2(t)\frac{\vel_{n_d}^2(t)}{c^2} \ra
\hfill (\ref{eq-en-gt3})
}$$
The last of Eqs.  (\ref{eq-en-gt3}) follows from the equality of the left side of Eqs. (\ref{eq-en-gt3}) with that of the time average of Eq. (\ref{eq-eng-totx2}b), $\la \g^2(t) \ra=1+ \la \g^2(t)\frac{\vel_{n_d}^2(t)}{c^2} \ra$. With Eqs. (\ref{eq-en-gt3}), $\la p_{n_d}^2(t)\ra=\la \g^2(t) M^2 \vel_{n_d}^2(t)\ra{}$ is written as 
$$\displaylines{\refstepcounter{equation} \label{eq-en-gt3.b} 
\hfill
\la p_{n_d}^2(t)\ra
= \la \g^2(t) \ra M \la \vel^2_{n_d}(t)\ra
\hfill (\ref{eq-en-gt3.b})
}$$
Combining with  (\ref{eq-en-gt3.b}), the root mean of  Eqs. (\ref{eq-eng-totx1}) is developed  as $\hbar \w_q-V_{n_d}= m c^2 =\sqrt{ M^2c^4 + \g^2 M^2  \vel^2_{n_d}  c^2 }$, where 
$$\displaylines{
\refstepcounter{equation} \label{eq-en-gtm}
\hfill
m = \sqrt{ \la m^2(t) \ra} =  \g M,  
\quad
  \g = \sqrt{ \la \g^2(t) \ra},
\quad
 \vel_{n_d}=\sqrt{\la \vel_{n_d}^2(t)\ra}. 
\hfill (\ref{eq-en-gtm})
}$$
Accordingly  the time averages of $\w_q(t)$, $p_{n_d}(t)$, etc. are given by $ \w_q= \sqrt{\la\w_q^2(t)\ra}=\g \W_q$, $p_{n_d}= \sqrt{\la p^2_{n_d}(t)\ra} = m \vel_{n_d} $, etc. Hence $\eng_{1_q}=\hbar \w_q$, where $\eng_{1_q}^0=\hbar \W_q$; $\eng_{n_d}=n \hbar \w_d= \frac{1}{2}m \vel_{n_d}^2+V_{n_d} = \g \eng_{n_d}^0$, where $\eng_{n_d}^0=n \hbar \W_d= \frac{1}{2}M \vel_{n_d}^2+V_{n_d}^0$. 
 
In sum, the dynamical variables $m, \w_q$ etc of the particle in harmonic, hence accelerated motion in the in $X$ direction here relative to  the laboratory frame $S$ are each on average augmented by the  factor $\g$  as measured in  $S$. Accordingly,  as measured in $S$, the space and time variables $x$, $\uscrx_{n_d}$, $\Acal_{n_d}$ and $t$ of the particle in  the particle's motion $X$ direction are each contracted by the average factor $\g$ from their rest values (indicated by the superscript $0$ as elsewhere if not  specified otherwise) $x^0$, $\uscrx_{n_d}^0$, $\Acal_{n_d}^0$ and $t^0$
as (Lorentz-Einstein transformation) $\frac{x}{x^0}=\frac{\uscrx_{n_d}}{\uscrx_{n_d}^0}=\frac{\Acal_{n_d}}{\Acal_{n_d}^0} =\frac{t}{t^0}=\frac{1}{\g}$. The space variables $z,\uscr_q$ etc. in the transverse direction  are particularly here the projections of the longitudinally contracted  (by $\g$) elastic deformation of the vacuum. These are therefore proportionally contracted  by the same factor $\g$  from their rest values:
$\frac{z}{z^0}=\frac{\uscrz_{1_q}}{\uscrz_{1_q}^0}=\frac{\Acal_{1_q}}{\Acal_{1_q}^0} =\frac{1}{\g}$. Finally,  $\Mcal_q =\g \Mcal^0_q$, $(\beta_d,\beta_q)=\g^3 (\beta_d^0,  \beta_q^0)$, $(\gw_d,\gw_q)=\g (\gw_d^0,\gw_q^0)$ given as derivative relations based on Eqs. (\ref{eq-Fqa.a}b), (\ref{eq-Fqa.b}b), (\ref{eq-engtotx1}) and (\ref{eq-engtotxd1}) combined with the foregoing transformation relations.

The author thanks Professor C Burdik for kindly inviting the author for  presenting a contribution at the $XXI$th Int Conf on Integrable Systems and Quantum Symmetries in  Prague, June, 2013,  
 during which stay  the author had very much enjoined extensive scientific discussions with several of the participants. 
The author's this research is privately financed by P-I Johansson (emeritus scientist, Uppsala Univ.).

\begin{appendix}

\section{Radiation damping based on solution to Maxwell equations 
}\label{app-damp}


\paragraph*{        
A.1 \ 
Electromagnetic Hamiltonian  quantisation and damping}  

For a self-contained illustration of the subject we derive
in this appendix  the electromagnetic equations  for $\engt_r $ (the radiation Hamiltonian), $\gw$, $F_r$, and $H_1$  based  on solutions to the Maxwell's equations.  The results are formally  mostly  well known;  although, the following solutions  will be obtained in terms of the radiation Hamiltonian, complex Poynting vector, and in the end, the quantised radiation fields. 

 Consider the quasi-stationary  charged oscillator as specified in Sec. \ref{Sec-2},  oscillating about position $\rb=0$ at time $t=0$ along  a specified $X_\i$ direction, which we set as the $Z$ axis here. In the case where the charged oscillator has also an (instantaneous) linear motion, and thus its radiated electromagnetic waves are  Doppler -differentiated  as in Sec. \ref{Sec-IEDWave}, we shall be concerned mainly with the superposed radiation electric field $\Efb(\rb,t)(=\Efb^\dagger(\rb,t)+\Efb^\ddagger(\rb,t))$ and  $\Bmbb(\rb,t)$ similarly as  in Sec. \ref{Sec-IEDWave}. Similarly 
as the $\Efb^j,\Bfb^j$, the $\Efb,\Bfb$ fields are governed by the  Maxwell's equations $\nablab \cdot \Efb =\frac{\rho_q}{ \ev_0} $, $ \nablab \times \Bfb = -\mu_0 \jb_q + \frac{1}{c'{}^2} \frac{\pd \Efb}{\pd t}$, $  \nablab \cdot \Bfb =0$,  $ \nablab\times \Efb= -\frac{\pd \Bb}{ \pd t}$, where  $c'{}^2=c^2 +\frac{V}{\Mv} $; $\rho_q$ and $  j_q$ are the charge and current densities of the radiation-emitting charged oscillator. In regions where $\rho_q=j_q=0$, choosing here the radiation  gauge $f= -\int \Phim dt$, $\Phim$ being the Coulomb potential of $q$,  such that $\nablab \cdot \Afb=0$ and (the vector potential) $\Afb$ will be a transverse field while $\Efb,\Bfb$ maintain unaltered, furthermore setting for simplicity $V=0$, the Maxwell's equations lead to the wave equations 
$$\displaylines{
\refstepcounter{equation} \label{wav-eqx}
\hfill
\nabla^2 \Yfb-\frac{1}{c{}^2}\frac{\pd^2 \Yfb}{\pd t^2}=0, \quad \Yfb=\Afb,  \Efb, \Bfb.
\hfill (\ref{wav-eqx})
}$$

Eqs. (\ref{wav-eqx}) have the solutions given in the spherical polar coordinates as, under the easily satisfied condition $\Acal/ r << 1 $ for the systems of interest here,
 $$\displaylines{
\refstepcounter{equation} \label{eq-Er}
\refstepcounter{equation} \label{eq-Era}
\hfill
\Afb  (\rb,t)               
=-  \frac{\ai^{1/2}\Av_0 r_0 \sin \theta  \sin (\kb\cdot \rb -\w t)       \hat{\theta}}{ r }, 
\quad \Av_0=\frac{q\Acal \w}{4\pi \ev_0 r_0 c^2 },
\hfill
(\ref{eq-Er})
\cr
\hfill \Embb(\rb,t) 
=-\nablab (\Phim + \frac{\pd f}{\pd t}) -\frac{\pd (\Avb -\nablab f) }{\pd t}
= -\frac{\pd \Avb  }{\pd t} 
=\frac{\ai^{1/2}\Emb_0 r_0 \sin \theta     \cos [\w (\frac{r}{c}-t)]   }{r } \ \hat{\theta} 
= \frac{ \Ef_{r\theta}\uscrzb_r}{\Acal}  \ \hat{\theta},
\hfill\cr   
\qquad\qquad\quad 
\hfill 
\cr 
\hfill
 \Ef_0=\Af_0 \w= \frac{q\Acal \w^2}{4\pi \ev_0 r_0 c^2 }, \quad 
\Ef_{r \theta}
=\frac{\Emb_0 r_0 \sin \theta   }{r }, \hfill
\cr
\hfill
\uscrzb_r(\rb,t)
= \ai^{1/2}\Acal \cos (\kb\cdot \rb -\w t)  \hat{ \theta}; \hfill(\ref{eq-Era}) 
\cr
\refstepcounter{equation} \label{eq-Erb}
\hfill \Bmbb=\nablab \times (\Avb-\nablab f)=- \frac{\Ef}{c} \hat{\phi};\hfill  (\ref{eq-Erb})
}$$    
 Here,  $\kb=k \hat {\rb}$, $k=\w/c $. $r_0 $ represents an (effective) radius of the oscillator. $\ai^{1/2}(t)\Emb_0 $  is the  radiation-damped amplitude of a spherical wave front at radius $r_0$ and  angle $\theta=\frac{\pi}{2}$ at time $t$, and  $\Emb_0$ is the amplitude without damping. $\ai$ is to be (separately) determined.   For either of the two oscillators of Sec. \ref{Sec-IEDWave} or any other compatible system, $\w=\g \W$,   $\Acal =\Acal^0/\g$, $\Mv=\g \Mv^0 $ and $r=({x^0}^2/\g^2+y{^0}^2/\g^2+z{^0}^2/\g^2)^{1/2}$  are the relativistic values of  $\W$, $\Acal^0$, $\Mv^0$  and $r^0$ as  in Sec. \ref{Sec-IEDWave}. The transverse coordinates $y,z$, representing the projections of longitudinally contracted vacuum for the same  (IED) particle here, are contracted similarly as  discussed in Sec. \ref{Sec-IEDtot}. Accordingly as derivative relations, $\Ef_0 =\g^2 \Ef_0^0$ and $\Efb =\g^2 \Efb^0$ in the above, and $\engt=\g \engt^0$,  $ \eng=\g \eng^0$,  $t_r=t_r^0/\g$ and  $L_r=L_r^0/\g$ later, are the relativistic values of $\Ef_0^0$, $\Efb^0$, $\engt^0$,  $\eng^0$, $t_r^0$ and  $L_r^0$.

$\uscrzb_r(\rb,t)$ of Eq. (\ref{eq-Era}) represents  the (quasi harmonic) oscillation displacement of a constituent of the vacuum, a coupled polarised  vacuuon according to \cite{jxzjied},  along $Z$ direction at position $\rb$ at time $t$, generated in response  to the perturbation by $\uscr(0,t-\frac{\rb}{c})$ of the charged oscillator at $\rb=0$ and  $t-\frac{\rb}{c}$. Accordingly, as follows directly from the last relation of Eqs. (\ref{eq-Era}a), $\Efb(\rb,t)$ is an electromagnetic representation of  this vacuum deformation  $ \uscrzb_r(\rb,t) $. The radiated time-dependent  electromagnetic Hamiltonian $\engt_r(r,\theta,t)$ therefore necessarily consists of  a kinetic oscillation term $ \engt_{r, kin} (r,\theta,t)$ and in addition, an elastic  vacuum  potential term  $\Vscr_{\v r}(r,\theta,t)$. The two terms and hence their sum  may be written down by  observing both the usual form of  electromagnetic energy equation and the underlying dynamics, given for per volume within a volume element $d \Vol$ about position $\rb$ at time $t$, as
$$\displaylines{
\refstepcounter{equation} \label{eq-Ekina}
\hfill
\frac{d \engt_{r, kin}}{ d \Vol}
=\frac{1}{2}  \left[\ev_0 \Embb^2(r,\theta, t, \mbox{\scriptsize{$-\frac{\pi}{2}$}} )
+ \frac{1}{\mu_0}  \Bmbb^2(r,\theta, t,  \mbox{\scriptsize{$-\frac{\pi}{2}$}} )\right]  
=\frac{1}{2}  \left[\ev_0 \frac{\dot{\Efb}^2(r,\theta, t)}{\w^2}
+ \frac{1}{\mu_0}  \frac{\dot{\Bfb}^2(r,\theta, t)}{\w^2}  \right]
\hfill 
\cr
\hfill
=\ev_0 \Ef_{r\theta}^2\ai \sin^2[\kb \cdot \rb -\w t]
=\Mcal^{em} \w^2 \Ef_{r\theta}^2\ai \sin^2[\kb\cdot \rb -\w t],
 \quad \Mcal^{em}=\frac{\ev_0}{\w^2}
\hfill(\ref{eq-Ekina})
\cr\refstepcounter{equation} \label{eq-Epota}
\hfill
\frac{d \Vscr_{\v r} }{ d\Vol}
=\frac{1}{2}  [\ev_0 \Embb^2(r,\theta, t)
+ \frac{1}{\mu_0}  \Bmbb^2(r,\theta, t)]  
=  \ev_0 \Efb^2(r,\theta, t) 
=  \ev_0 \Ef_{r\theta}^2\ai \cos^2[\kb\cdot \rb -\w t ]
\hfill (\ref{eq-Epota})
\cr
\refstepcounter{equation} \label{eq-rhoeng}
\hfill
\frac{d \engt_r}{ d \Voll}
=\frac{d ( \engt_{r,kin}+ \Vscr_{\v r})        }{        d \Voll } 
=
\ev_0\Ef_{r \theta}^2
a_1 [\sin^2 (\kb\cdot \rb-\w t)+\cos^2 (\kb\cdot \rb-\w t)]
 =  \frac{ \ev_0 r_0^2\sin^2 \theta \  |\Efb_x^c(\rb,t)|^2}{r^2}, \hfill
\cr \hfill(\ref{eq-rhoeng})
\refstepcounter{equation} \label{eq-rhoeng-wave}
\cr
\hfill
\quad  \Efb_x^c(\rb,t)=\ai^{1/2} \Ef_0 e^{-i (\kb \cdot \rb-\w  t    )}  \hat{\theta}
=\ai^{1/2}\Ef_0 \psi_r(\rb,t) \hat{\theta}=   \frac{\Ef_0}{\Acal} \uscrz_r^c(\rb,t) \hat{\theta}, 
\hfill\cr
\hfill
\psi_r(\rb,t)= e^{-i (\kb \cdot \rb-\w  t    )}, \quad 
\uscrz_r^c(\rb,t)=\ai^{1/2}  \Acal  \psi_r(\rb,t)
 \hfill (\ref{eq-rhoeng-wave})
}$$ 
In the above, the first of Eqs. (\ref{eq-Ekina}a)  preserves the usual form of electromagnetic energy equation except with the cosine function being  shifted by a phase $-\frac{\pi}{2}$. The second of Eqs. (\ref{eq-Ekina}a)  is alternatively and equivalently expressed as the time rate of $\Ef$; that is, $\engt_{kin} $ is now proportional to $\dot{\Ef}^2$ and a corresponding "electromagnetic inertial mass" $\Mcal^{em}$, by making analogy to the kinetic  energy of an oscillator. It is rather the elastic potential term (\ref{eq-Epota}), which does not present in the usual empty  vacuum representation, 
 that formally directly corresponds to the usual (phenomenological) electromagnetic energy equation.

The "complex Poynting vector", defined  as  the vector intensity of the electromagnetic radiation  Hamiltonian $d \engt_r(r,\theta,t)$  passing per unit time per unit area through a differential cross-section area $d \sig =\frac{d \Vols}{dr}=r^2 \sin \theta d \theta d \phi $ along $\rb (r,\theta, \phi)$ direction, is given as 
$$\displaylines{
\refstepcounter{equation} \label{eq-Z1}
\hfill
\Ib^c(r,\theta,t)=\frac{d\engt_r(r,\theta,t) }{d t d \sig}\ \hat{r}
= \frac{d \engt_r(r,\theta,t)}{ d \Vol}c   \ \hat{r}
=  \frac{  \ev_0 r_0^2\sin^2 \theta \ |\Efb_x^c(r,t)|^2 c }{r^2}   \  \hat{r}
\hfill
 (\ref{eq-Z1})
}$$ 
The negative time rate of the Hamiltonian of the oscillator, $-\frac{d \engt(t)}{dt }$, is equal to the total radiation power passing a sphere of radius $r$, i.e.,  $\frac{d \engt_r(r,\theta,t)}{d t d \sig} d \sig=I(r,\theta,t) d \sig$ integrated over $4\pi $ solid angle: 
$$\displaylines{
\refstepcounter{equation} \label{eq-Pe}
\hfill 
-\frac{d \engt(t)}{dt }=P(t)\equiv \frac{d \engt_r(t)}{dt}
= \int_0^\pi\int_0^{2\pi} I(r,\theta,t) r^2 \sin \theta d \theta d \phi 
= s_0  \ev_0 |\Efb_x^c(r,t)|^2  c    
\quad \hfill (a) \quad 
\cr
\qquad  \qquad\qquad \qquad \qquad \quad 
=\frac{q^2\w^4 \Acal^2 \ai}{6\pi \ev_0c^3} 
 =\frac{q^2 |\ddot{\uscrz}_r^c{}|^{2}}{6\pi \ev_0 c^3}
 \hfill  (b) \quad 
\cr
\qquad\qquad \qquad \qquad \qquad \quad 
= \frac{q^2 \w^4  \Acal^2\ai \frac{1}{2}\Mv }{6\pi \ev_0c^3\frac{1}{2}\Mv} 
=\frac{q^2\w^2\engt_\mus}{3\pi \ev_0 c^3 \Mv} 
= \gwsm_\mus \engt, 
\quad \hfill (c) \quad
\cr
\hfill (\ref{eq-Pe}) 
\cr
\refstepcounter{equation} \label{eq-Pe3}
\hfill
 \gwsm_\mus
=\frac{q^2\w_\mus^2}{3\pi \ev_0 c^3 \Mv}
=\frac{q^2 \w^3 \Acal_1^2}{6 \pi\ev_0 c^3 \hbar}
=\frac{4 q r_0 \Efb_0 }{3 \Acal \Mv c  }
\hfill(\ref{eq-Pe3})                          
}$$
where $ s_0=\frac{8\pi }{3}r_0^2$ represents the apparent  area passed by the spherical wave  $\Efb$ at $r=r_0$. Eqs. (\ref{eq-Pe}c) are given after multiplying and then dividing the right sides of (\ref{eq-Pe}.b) by $\frac{1}{2}\Mv$, and substituting into it by $\engt= \frac{1}{2}\Mv\w^2\Acal ^2 {} \ai(t) $  given by  (\ref{eq-engnx2pa}a). The second expression of Eqs. (\ref{eq-Pe3})  is given by substituting Eq.  (\ref{eq-engnx2pa-p}b) for $\Mv$; and the  third by substituting (\ref{eq-Era}a)  for $\Ef_0$. Eqs. (\ref{eq-Pe}b) and (\ref{eq-Pe3}) give the well-known  Larmor formula   in complex form here    and  formula for damping factor. Integrating (\ref{eq-Pe}c,a) over time  $(0,t)$ gives $\ln  \engt_{\mus}(t)|^t_0=-\gwsm_\mus t |^t_0$, $-  \engt(t) |^t_0 =\engt_r(t)|^t_0$. Combining with the quantisation results of (\ref{eq-engnx2pa-p}a)--(b) for the same energies $\engt$, $\eng$ as  here, we obtain
$$\displaylines{
\refstepcounter{equation} \label{eq-engrate3}
\hfill
\engt(t)  \rightarrow \engt_n(t)
= \eng_n e^{-\gw_\mus t} = \eng_n\ai(t), \quad 
\eng_n =\engt_n(0)
=\frac{1}{2} \Mv \w^2 \Acal_n^2, \quad      
\ai(t)=e^{-\gw t}
\hfill (\ref{eq-engrate3})
\cr
\refstepcounter{equation} \label{eq-engrate3b}
\hfill 
\engt_{r}(t) \rightarrow \engt_{r,n} (t)
=\eng_{r,n} (1- e^{-\gw t})
=\eng_{r,n}\aii(t)
\ \  
\eng_{r,n} =\engt_{r,n} (\infty)
=L_r s_0\ev_0 \Ef_{0 n}^2, 
\ \  \aii(t)=1-e^{-\gw t},
\hfill
\cr
\hfill
{\rm where} \quad 
L_{r} =\frac{1}{\a} c =t_{r} c= \frac{3\pi \ev_0 c^4 \Mv}{q^2 \w^2}, \quad t_r=\frac{1}{\gw},
\quad 
\Ef_{0 n}= \frac{q\Acal_n \w^2}{4\pi \ev_0 r_0 c^2 }. 
\hfill      (\ref{eq-engrate3b})
}$$
The equations for $\ai,\aii$  above are the same as  Eqs.  (\ref{eq-engnx2pa-p}c), (\ref{eq-engnx2pp1-A}b). $t_r$ of (\ref{eq-engrate3b}) is identical  to the mean transition time $ \la t\ra=  \frac{\int^\infty_0 t e^{-\gw t} d t }{ \int^\infty_0  e^{-\gw t} d t} =\frac{1}{\gw} $. So $L_{r}$ represents an average  length of the radiated electromagnetic wave train. 

Specifically for a charge oscillator $q=+e$ or $-e$ which oscillates in the vacuum potential field  $V_{\v q}$ and generates an electron or proton based on Eqs. \ref{eq-Vvq}--\ref{eq-contnu3-a.b}, Sec. \ref{Sec-5}, accordingly with the substitutions of $ | \pm e|$, $\Mcal_q=\frac{2\hbar }{\Acal_{1_q}^2 \w_q}=\frac{2\hbar^2}{\Acal_{1_q}^2 mc^2}$ (given after Eq.   \ref{eq-engnx2pa-p}d), $\w_q=\frac{m c^2}{ \hbar} $ for $q$, $\Mv$, $\w$, Eqs. (\ref{eq-Pe3}), (\ref{eq-engrate3b}b), (a) and (c)  are written as 
$$\displaylines{
\refstepcounter{equation} \label{eq-gwq}
\hfill
\gw_q 
=\frac{e^2\w^2_q }{3\pi \ev_0 c^3 \Mcal_q}
= \frac{e^2 (\frac{mc^2}{\hbar})^2    }{3\pi \ev_0 c^3 (\frac{2\hbar^2}{\Acal_{1_q}^2 mc^2})}
= \frac{e^2 m^3 c^3 \Acal_{1_q}^2   }{6\pi\ev_0 \hbar^4}, 
\quad
t_{r,q}=\frac{1}{\gw_q}, \quad
L_{r,q}=c t_{r,q}=\frac{6\pi \ev_0 \hbar^4}{e^2 m^3 c^2 \Acal_{1_q}^2},
\hfill
\cr
\hfill
 \Ef_{0 1_q}=\frac{e \Acal_{1q} \w_q^2}{4\pi\ev_0 r_0 c^2}=\frac{e\Acal_{1_q}m^2 c^2}{4\pi\ev_0 r_0\hbar^2}
\hfill
(\ref{eq-gwq})
}$$
Calculated values of $\gw_q$, $t_{r,q}$, $L_{r,q}$ and $E_{0,1_q}$    based on Eqs. (\ref{eq-gwq}) for the charge oscillators $-e $ and $+e$,  which generate an IED electron and  proton, are given in Table \ref{tab1}, where the $\Mcal_q$ and $\w_q(=\frac{m c^2}{\hbar})$ values (second and third columns in the table)
are as input data. 
\begin{table}[here]
{\footnotesize            
\vspace{-0.5cm}

\begin{center} 
\caption{ } 
\label{tab1} 
\begin{tabular}[h]{
|p{1.2cm}  
 |p{1.8cm}         
 |p{1.6cm}     
|p{1.6cm}        
|p{1.7cm} 
 |p{1.6cm} 
|p{1.5cm}  |  
}
\hline
      Oscillator      
& $\hspace{0.5cm}\Mcal_q^{(a)} $           
& \mbox{$\w_q(=\frac{m c^2}{\hbar})$}    
& \mbox{\hspace{0.6cm}$\a_q$ }             
& \hspace{0.6cm}$t_{r,q}$                         
&\hspace{0.4cm} $L_{r,q}$           
& \hspace{0.6cm}$\Ef_{01_q}^{(b)}$               
\\
\hspace{0.4cm}$q:$          
& \hspace{0.5cm}(kg)   
& \hspace{0.4cm}(r/s)      
&\hspace{0.4cm} (r/s)                
& \hspace{0.6cm}(s)&
 \hspace{0.6cm}(m)         
&\hspace{0.4cm} (N/C)     
\\ 
\hline
 \hspace{0.3cm}$-e$                                                  
&\mbox{$1.083\times{}10^{-18}$}         
& \mbox{$7.763 \times 10^{20}$}      
& \mbox{$6.354 \times 10^{6}$} 
& \mbox{$1.574\times 10^{-7}$} 
& \mbox{$47.18$}           
& \mbox{$9.656\times 10^{15}$}           
\\ \hline
 \hspace{0.3cm}$+e$                                                  
&\mbox{$5.896\times 10^{-22} $}         
& \mbox{$1.425 \times 10^{24}$}      
& \mbox{$3.935 \times 10^{16} $} 
& \mbox{$2.541\times 10^{-17}$} 
& \mbox{$7.619\times 10^{-9}$}  
& \mbox{$3.256\times 10^{22} $}          
\\ \hline
\end{tabular}

\begin{minipage}[h]{13.cm}\baselineskip 0.4cm
\footnotesize{\baselineskip 0.4cm
$(a)$: Values from  [\citejxzjied b], with the free parameter $f_1$ set  to 1.
\\
$(b)$: Based on Eq. (\ref{eq-gwq}b), where we have set $r_0=\Acal_{1q}$; clearly  $\Acal_{1q}\sim \frac{b_\v}{2}$, $b_\v \sim 1 \times 10^{-18}$ m being  the inter-vacuuon distance estimated based on  experiment [\citejxzjied b].  
}
\end{minipage}
\end{center}
}
\end{table}
In comparison, for an electron oscillator  described by Eqs. (\ref{eq-Fqa.a})--(\ref{eq-contnu3-a.a}), Sec. \ref{Sec-5},  with an oscillation frequency $\w_d=2\pi \cdot 10^{14}$ r/s, corresponding evaluations give $t_{r,e}= 2.02 \times 10^{-7}$ s, $L_{r,e}=60.6$; and for a proton oscillator with the same oscillation frequency, $t_{r,p}=3.71\times 10^{-7}$ s, $L_{r,p}=1.11 \times 10^5$ m. Common with  all of the examples, the condition $\gw << \w$ is well satisfied.


\refstepcounter{subsection}\label{app-damp.2}
\paragraph{
A.2 \ 
Radiation--charge  electromagnetic interaction force and work}  

In the mechanical representation of Sec. \ref{Sec-2} the radiation damping force $F_r$ acts on a charged oscillator through the deformation 
 of a  viscous elastic vacuum medium, $\uscrz_{r} \propto  \uscrz \propto \Acal$ whose  electromagnetic counterpart is   $\Ef$ (Eq. \ref{eq-Era}). We shall below derive the electromagnetic counterpart  of the $F_r$, $F_r^{em}$. Consider that a charged oscillator  $\mu$ of charge $q$ is emitting radiation of radiation electric  field $\Efb(r,\theta, t)$ at position $r$ at time $t$; our specific attention is the  $r$ lying  in the region $[-r_0,r_0]$ occupied by the oscillator here. $\mu$ must  inevitably in turn be acted by $\Efb$ a Coulomb force given for per unit length along the $X$ axis and per unit cross-sectional area as, with (\ref{eq-Er}) for $\Efb$, 
$$\displaylines{
\refstepcounter{equation} \label{eq-Fb}
\hfill
\Fb_e(x,t)
= -q \Efb (r,\frac{\pi}{2},t)
=- \frac{\ai^{1/2} q^2\Acal \w^2\cos [k x -w t]  \hat{z} }{4\pi\ev_0 r c^2}  
\qquad\qquad
 \hfill
\cr
\hfill
=- \Mv\left(\frac{q^2\w^2}{3\pi\ev_0  c^3\Mv}\right) \frac{3 \ai^{1/2} c \Acal \cos [\w(\frac{x}{c}-t)] \hat{z} }{4r}
=- \frac{3 \gw_\mus  \Mv   c \uscrzb(x,\frac{\pi}{2},t) }{4r}
\hfill (\ref{eq-Fb})
}$$ 
For obtaining the last of Eqs.  (\ref{eq-Fb}) we used  (\ref{eq-Pe3}) and $\uscrz(t)$ for $ \uscrz_r(x, \frac{\pi}{2}, t)=\ai^{1/2} \Acal \cos [\w(\frac{x}{c}-t)]  $.  

Substituting into the last of Eqs. (\ref{eq-Fb}) the identity relation $\uscrz (t)=\frac{d \uscrz (t')}{\w d t} $ (where $t'=t-\frac{\pi }{2\w}$), and $r=r_0$ and $\w = 2\pi c/\lam$, we obtain
$$\displaylines{
\refstepcounter{equation} \label{eq-Fex}
\hfill
 \frac{s_0 }{r_0 \lam}\Fb_e
=-\gw_\mus \Mv    \frac{d \uscrzb_\mus}{d t}
\hfill (\ref{eq-Fex})
}$$ 
where $s_0=\frac{8\pi }{3}r_0^2$ as before. The right side of (\ref{eq-Fex}) is just the expression for $\Fb_r (=-\gw \Mv \frac{d\uscrx}{d t})$ of  Sec. \ref{Sec-2}, hence  
$$\displaylines{
\refstepcounter{equation} \label{eq-Fb2}
\hfill
       \Fb_{r}^{em}(=\Fb_r )           
=\frac{s_0 \Fb_e }{r_0 \lam} 
= - \frac{s_0 q\Efb }{r_0 \lam} 
= - \frac{s_0 q\w\Afb  }{r_0 \lam} 
\hfill (\ref{eq-Fb2})
}$$

We shall next derive the source-radiation interaction potential ($V_{\mus \Isub}$) from the work done by $F_r$, or $F_r^{em}$. $F_r^{em}$ does to the oscillator $\mu$ a dissipative work $W_\Isub$; and $F_e$, hence $F_{r}^{em}$ or $F_r$, is a conservative force (i.e., it depends on the position of the charge in the $\Efb$ field only). So $W_\Isub$ amounts to a negative source-radiation interaction potential, $V_{\mus \Isub}$. Suppose that at time $t=0$, the oscillator has a potential energy $V(t=0)=V_0 =\frac{1}{2}\Mv \w^2 \bar{\uscrz}^2$ ($V(t)$ is a short hand notion of $V(\uscr(t))$ used in this and the next sub-section), with $\bar{\uscrz}=\frac{\uscrz}{\ai^{1/2}}$, which is undamped. The corresponding work done during  one oscillation cycle 
$2\pi \tau /\tau=2\pi$ int  $0\le  t'\le \tau$,  thus is given as, with $d \uscrz/dt = -\w \uscrz(x,t -\pi/2\w) $, 
$$\displaylines{\refstepcounter{equation} \label{eq-Vqr}
\hfill\qquad
 \D V_{\mus \Isub}(0)
=-4 \times  \frac{2\pi (\tau/4)}{\tau}  \int_{\uscrz(0) =0} ^{\uscrz(\frac{\tau}{4})=\Acal_{n} } 
F_{r} (t')d \uscrz (t') 
=2\pi \int_{\uscrz(0) =0}^{\uscrz(\frac{\tau}{4})=\Acal_{n} }  
\Mv \gw_\mus \w  \uscrz (t')d \uscrz(t')  \qquad \quad \hfill
\cr\qquad
\qquad \qquad 
= - (\frac{1}{2} \Mv \w^2  \uscrz^2(0)) 2\pi \frac{\gw_\mus  \tau }{2\pi}  
 =-V_{0}  \gw_\mus \tau  
\dot{=} - V_0 (1-e^{-\gw_\mus \tau}) \hfill (\ref{eq-Vqr})
}$$
The last equality holds  for $\gw_\mus \tau <<1$, a  condition characteristic  with  the quasi harmonic oscillation here. The total potential at time $t=\tau$ 
thus is, 
$
          V(\tau)=V(0)+\D V_{\mus \Isub}(\tau)= V_0- V_0 (1-e^{-\gw_\rr \tau})
         =V_0 e^{-\gw_\mus \tau} 
$.
Similarly, for $\mbox{$\tau \le t'\le 2\tau$} $,
$\D V_{\mus \Isub}(2\tau)= -V(\tau) (1-e^{-\gw_\mus \tau})$, $  V_{{}}(2\tau)=V_{{}}(\tau)+\D V_{\mus \Isub}(\tau) =V_{{}}(\tau) e^{-\gw_\mus \tau}=  (V_0 e^{-\gw_\mus \tau})e^{-\gw_\mus \tau}       = V_0 e^{-\gw_\mus 2\tau}$; and so forth. Finally, for  $t-\tau \le t' \le t$, $\D V_{\mus \Isub}(t)= -V(t-\tau) (1-e^{-\gw_\mus \tau})$,
$$\displaylines{
\refstepcounter{equation} \label{eq-Vq-rqa}
\hfill          
V_{{}}(t)=V_{{}}(t-\tau)+\D V_{\mus \Isub}(t)
= (V_0 e^{-\gw_\mus (t-\tau)}) e^{-\gw_\mus \tau}
= V_0 e^{-\gw_\mus t} 
= V_0 + V_{\mus \Isub}(t) 
\hfill 
\cr
\hfill
H_\Isub 
=V_{\mus \Isub}(t)
= -V_0 (1-  e^{-\gw_\mus t}) 
= -\frac{1}{2} \Mv_\mus  \w_\mus^2 x_\i^2(1-e^{-\gw_\mus t})
=-\hbar \w_\mus \cos^2 (\w_\mus t) (1-  e^{-\gw t})
\hfill (\ref{eq-Vq-rqa})
}$$
Eqs. (\ref{eq-Vq-rqa}) and $H_0=H-H_\Isub$ are the same as  Eqs. (\ref{eq-contnu3-a}.b),  (\ref{eq-Xim2}c), and (\ref{eq-Xim2}a-b), Sec. \ref{Sec-SHO-Quat}. 

\refstepcounter{subsection}\label{app-damp.3}

\paragraph{A.3 \ Radiation-charge interaction Hamiltonian expressed in $\Afb$}  Suppose that due to the action of  the $V_{\mus \Isub}(t)$ above, a photon is emitted by its source  during a particular  transition time $t_r =\tau_\mus=2\pi/\w_\mus$, and we want to express $H_\Isub=V_{\mus \Isub}$ using $\Afb$. For $\gw_\mus t_r = \gw_\mus \tau_\mus <<1$, Eq. (\ref{eq-Vq-rqa}b) reduces to
$$\displaylines{
\refstepcounter{equation} \label{eq-H1b}    
\hfill
  H_\Isub= V_{\mus \Isub}(\tau_\mus) 
= -\hbar \w_\mus \cos^2 (\w_\mus \tau_\mus) (1-(1-\gw_\mus \tau_\mus +\frac{1}{2!}\gw_\mus^2 \tau_\mus^2 -\ldots ) )
\dot{=} - 2\pi \hbar \gw_\mus 
\hfill
\cr
\hfill
{\rm or} \quad H_\Isub
= -2\pi \hbar (\frac{4qr_0 \Af_{} \cdot (\vel k)}{3\Acal \Mv c})
= -  \frac{ \hbar q (\frac{8\pi  r_0 \vel }{3 \Acal c} ) \Af  (-i \nabla_{x_j}) }{ \Mv_\mus } \hfill (\ref{eq-H1b})
}$$
The first of Eqs. (\ref{eq-H1b}) is given after substituting   (\ref{eq-Pe3}) for $\a$,  $\Ef_0 =\Af_0\w \approx \Af \w$   from (\ref{eq-Era}a) and  $\w=\vel k$; and the second $k=-i \nabla_{x_j}$.   If the  radiation is emitted by the charge oscillator $q$ along  the $Z$ direction as specified in Sec. \ref{Sec-5},  substituting the corresponding variables $z$, $\Mcal_q$, $\Acal_q$, $\w_q=kc$ and $c$ for $x_j$, $\Mv$, $\Acal$,  $\w$ and  $\vel$, (\ref{eq-H1b}) is written as
$$\displaylines{
\refstepcounter{equation} \label{eq-H1b.a}    
\hfill
H_\Isub
=\frac{i\hbar q (\frac{8\pi r_0 }{3 \Acal_q}) \Afb\cdot \nablab_{z}}{\Mcal_q}
= \frac{i\hbar q  \Afb'\cdot \nablab_{z}}{\Mcal_q},
\quad
\Afb' =(\frac{8\pi r_0 }{3 \Acal_q}) \Afb 
\hfill (\ref{eq-H1b.a})
} $$
If alternatively the radiation is emitted by  the charged particle oscillator $d$ along the $X$ direction, substituting accordingly $x$, $m$, $\Acal_d$, $\w_d=\frac{1}{2}\vel_{n_d} k_d$,  $\frac{1}{2}\vel_{n_d}$, and $k_d$ for $x_j$, $\Mv$, $\Acal$,  $\w$,  $\vel$, and $k$, (\ref{eq-H1b}) is written as
$$\displaylines{
\hfill
H_\Isub =   \frac{i \hbar q (\frac{8\pi  r_0 \frac{1}{2}\vel_{n_d} \hat{y}  }{3 \Acal_d c} \times \Afb) \cdot   \nablab_x }{ m }
=   \frac{i \hbar q (\Pb_r \times  \Afb) \cdot   \nablab_x }{ m }
=   \frac{i \hbar q \Afb\pa \cdot   \nablab_x }{ m }, 
\hfill 
\cr
\refstepcounter{equation} \label{eq-H1c}    
\hfill  
\Ab\pa= \Pb_r \times \Afb, 
 \quad \Pb_r
=   \frac{8\pi  (r_0/c) }{6 (\Acal_d/\vel_{n_d} ) }  \hat{y}
\hfill (\ref{eq-H1c})
}$$
The electromagnetic radiation fields, the vector field $\Afb$ in the above,  emitted (or absorbed) by either the oscillator $q$ or $d$  is oriented along the transverse $Z$ direction and propagated in the $X$ direction. So when expressed into the dot product form in the second to third  of Eqs.  (\ref{eq-H1c}), an apparent $\Afb''$ is involved given after the projection of  $\Afb$ by  $\Pb_r$.

The same result (\ref{eq-H1c}) for  a charged particle oscillator $d$ is in the usual derivation given by considering the effect of $\Afb$ and  $\Phim$ on the particle's linear momentum. Assume that $\Afb$  is along the transverse $Z$ direction as in the above and is projected on to  the $X$ direction into $\Afb\pa$  according to Eq. (\ref{eq-H1c}). Under the action of $\Afb''$, the  linear momentum of the charged oscillator is reduced from $\pb_d$ 
 by an amount  $\D \pb_d= -q \Afb\pa$ to $\pb'_d =\pb_d + \D \pb_d=\pb_d -q \Afb\pa$.  The Hamiltonian operator $H_{d}$, subtracted by $V_{d0}$, is thus given as, with the substitutions $ \pb_d \cdot (q \Ab'') = \Afb\pa \cdot \pb_d -i \hbar \nablab\cdot \Afb\pa$ and $\pb_d = \frac{\hbar }{i} \nablab$, 
$$\displaylines{
\refstepcounter{equation} \label{eq-EbH1}
\hfill 
H_d-V_{d0}= \frac{1}{2m} (\pb_d - q \Afb\pa)^2+ q\Phim    
=  \frac{1}{2m}  [ \pb_d \cdot \pb_d -2q\Afb\pa \cdot \pb_d
      + i\hbar\nablab \cdot \Afb\pa
 +q^2 \Afb\pa^2 ] +q\Phim   \hfill
\cr
\hfill
\quad 
= -\frac{\hbar ^2}{2m}\nabla^2+ \frac{i \hbar q\Afb\pa \cdot \nablab}{m}  
+ \frac{i\hbar q \nablab\cdot  \Afb\pa}{2m} + \frac{q^2 \Afb\pa{}^2}{2m} + q \Phim 
\dot{=} -\frac{\hbar ^2}{2m}\nabla^2+ \frac{i \hbar q\Afb\pa \cdot \nablab}{m} \hfill
(\ref{eq-EbH1})
} $$
The last of (\ref{eq-EbH1}) is given by considering the situation that $ \nablab\cdot  \Afb\pa=0$ for $\Afb\pa$ is essentially of uniform amplitude along the radiation path;  $ \frac{q^2 \Afb\pa{}^2}{2m} $ is a higher order term and thus omitted;  and the static $\Phim $ field produced by the charge $q$ does not act on $q$ itself, thus $q\Phim=0$. The remaining  last term in the last of Eqs. (\ref{eq-EbH1}) gives just the $H_{d \Isub}$.


\end{appendix}

\end{document}